\documentclass[aps,prx,reprint,amsmath,amssymb,floatfix,superscriptaddress]{revtex4-2}

\usepackage{multirow}
\usepackage{booktabs}
\usepackage{graphicx}
\usepackage{dcolumn}
\usepackage{bm}
\usepackage[dvipsnames,table]{xcolor}
\usepackage{gensymb}
\usepackage[normalem]{ulem}
\usepackage{float}
\usepackage{textcomp}
\usepackage[colorlinks=true,linkcolor=blue,citecolor=blue,urlcolor=blue]{hyperref}

\newcommand{\graymid}{\arrayrulecolor{gray!30}\midrule\arrayrulecolor{black}}
\newcommand{\ket}{\rangle}
\newcommand{\bra}{\langle}
\newcommand{\sixJ}[6] {\left\{\begin{array}{ccc} #1 & #2 & #3 \\ #4 & #5 & #6 \end{array}\right \}}
\newcommand{\threeJ}[6] {\left(\begin{array}{ccc} #1 & #3 & #5 \\ #2 & #4 & #6 \end{array}\right )}
\newcommand{\nineJ}[9] {\left\{\begin{array}{ccc} #1 & #2 & #3 \\ #4 & #5 & #6 \\#7 & #8 & #9 \end{array}\right \}}

\newcommand{\p}{\prime}
\newcommand{\pp}{{\prime\prime}}

\begin{document}

\title{Probing $P,T$-Symmetry Violation with Optically Trapped Asymmetric Top Molecules}

\date{\today}
\author{Yuxi Yang}
\thanks{Contact author: yuxiyang@caltech.edu and arianjad@mit.edu; Y.Y.\ and A.J.\ contributed equally to this work.}
\affiliation{Division of Physics, Mathematics, and Astronomy, California Institute of Technology, Pasadena, USA}

\author{Arian Jadbabaie}
\thanks{Contact author: yuxiyang@caltech.edu and arianjad@mit.edu; Y.Y.\ and A.J.\ contributed equally to this work.}
\affiliation{Harvard/MIT Center for Ultracold Atoms, Cambridge, MA, USA}
\affiliation{Department of Physics, Massachusetts Institute of Technology, Cambridge, USA}
\affiliation{Department of Physics, Harvard University, Cambridge, USA}

\author{Lukáš Félix Pašteka} 
\affiliation{Van Swinderen Institute for Particle Physics and Gravity, University of Groningen, Groningen, The Netherlands}
\affiliation{Nikhef, National Institute for Subatomic Physics, Amsterdam, The Netherlands}
\affiliation{Department of Physical and Theoretical Chemistry, Faculty of Natural Sciences, Comenius University, Bratislava, Slovakia}

\author{I.~Agustín Aucar}
\affiliation{Van Swinderen Institute for Particle Physics and Gravity, University of Groningen, Groningen, The Netherlands}
\affiliation{Nikhef, National Institute for Subatomic Physics, Amsterdam, The Netherlands}
\affiliation{Instituto de Modelado e Innovación Tecnológica (IMIT), CONICET--Universidad Nacional del Nordeste, Corrientes, Argentina}

\author{Rob G. E. Timmermans}
\affiliation{Van Swinderen Institute for Particle Physics and Gravity, University of Groningen, Groningen, The Netherlands}
\affiliation{Nikhef, National Institute for Subatomic Physics, Amsterdam, The Netherlands}

\author{Nicholas R. Hutzler}
\affiliation{Division of Physics, Mathematics, and Astronomy, California Institute of Technology, Pasadena, USA}

\begin{abstract}
Searches for the $P,T$-violating electromagnetic moments are among the most sensitive probes of symmetry-violating physics beyond the Standard Model (BSM). Extending beyond the current limits will benefit from molecules that have fully controllable orientation at low electric fields, retain long coherence times, and are laser-coolable -- all of which are offered by asymmetric top molecules (ATMs). By exploiting the intrinsic rotational $K$-doubling in ATMs and the associated long-lived ($T_1 \gg 10$~s) parity doublets afforded by $C_{2v}$ symmetry and nuclear-spin statistics, these species combine large electric polarizability with long coherence times in modest laboratory fields ($\lesssim 1$~kV/cm). We study alkaline-earth(-like) monoamides, $\mathcal{M}$--NH$_2$ ($\mathcal{M}$ = Ca, Sr, Ba, Yb, Ra), which possess favorable electronic structure for laser cooling. To enable accurate modeling, we perform \textit{ab-initio} calculations of fine and hyperfine constants, identifying the importance of relativistic effects in determining the spin-rotation tensor. These parameters are used with an effective Hamiltonian approach to model the rotational and hyperfine structure of the vibronic ground state, quantifying electron electric dipole moment (EDM) sensitivities and identifying feasible measurement schemes. We compute the effect of external fields and identify engineered clock transitions that suppress sensitivity to external perturbations while retaining strong EDM sensitivity. We further characterize the magic trapping conditions that null differential light shifts in an optical trap and maintain coherence. Under these conditions, we project a statistical electron-EDM sensitivity over an order of magnitude beyond the reach of current best experimental limits, with further gains available from increased molecule number and coherence time. Our results establish asymmetric top molecules as viable species with a tunable quantum-state structure that can enable sensitive molecular symmetry violation measurements with long coherence times in the search for BSM physics.
\end{abstract}

\maketitle

\section{Introduction}
Searching for Charge-Parity ($CP$) violating physical observables, e.g., a permanent electric dipole moment (EDM) of a fundamental particle, is a promising approach to answering open questions in fundamental physics, such as the origin of the observed imbalance of matter over antimatter in the universe \cite{Sakharov1967,Pospelov2005}. A permanent EDM violates both parity ($P$) and time-reversal ($T$) symmetry and, via the $CPT$ theorem, probes $CP$ violation. Due to small Standard Model EDM contributions, experimental detection of an EDM near current state-of-the-art sensitivities would indicate new physics beyond the Standard Model (BSM). Many platforms have been used for successful EDM measurements, including neutrons, nuclei, atoms, and molecules~\cite{GingesViolations2004,EngelElectric2013,Safronova2018,ChuppElectric2019,AlarconElectric2022}.

Molecular experiments currently lead these searches: measurements with the diatomic molecules HfF$^+$ and ThO have pushed the reach of new-physics detection to the $\sim$10~TeV scale~\cite{Roussyimproved2023,ACMECollaborationImproved2018}. Next-generation searches with neutral species will strongly benefit from laser cooling for high-fidelity state preparation, readout, and trap loading~\cite{Shuman2010,TruppeMolecules2017,Anderegg2019}, together with the long coherence times available in conservative traps~\cite{Carr2009,Baranov2012}. In heavy diatomics, however, the parity-doublet structure desirable for EDM measurements interferes with the electronic structure required for photon cycling. Polyatomic molecules resolve this conflict~\cite{Kozyryev2017PolyEDM}: their additional mechanical degrees of freedom generically supply parity doublets, while an alkaline-earth metal bonded to an electronegative ligand keeps the valence electron in an anti-bonding orbital localized away from the bond, preserving the cycling transitions~\cite{IsaevPoly,KozyryevPolyCool,liEmulatingOpticalCycling2019,Ivanovrational2019,Isaev_Ra}.

Parity doublets are central to modern EDM measurement schemes. With a molecular-frame (symmetry-allowed) dipole moment $D_0$ and a parity splitting $\Delta E_\pm$, electric fields of order $\mathcal{E}\simeq 2~\text{V/cm}\times\Delta E_\pm / D_0 \times \text{Debye} / (h \, \text{MHz})$ are sufficient to realize $\mathcal{O}(1)$ polarization~\cite{HutzlerPolyatomic2020}. When polarized, the molecule's low- and high-field-seeking states are oppositely oriented, forming internal co-magnetometer states that allow spectroscopic reversal of the EDM interaction without modifying laboratory fields~\cite{Kozyryev2017PolyEDM}.

These advantages motivated successful efforts~\cite{vilas_magneto-optical_2022,AndereggQuantum2023} to produce, cool, and trap linear triatomic molecules, using the excited vibrational bending mode in the ground electronic state as the science state with $\Delta E _\pm \sim  30~\text{MHz}$ parity doublets. In particular, applications of quantum state engineering to linear triatomic molecules have achieved significant improvements in robustness against noise and systematics, without compromising EDM sensitivity~\cite{TakahashiEngineering2023,AndereggQuantum2023,Bause2025}. However, the science states used in the EDM measurement are excited vibrational states, with radiative lifetimes $T_1\sim500~\text{ms}$~\cite{HallasOptical2023,Bause2025}. This not only limits the time available for a measurement, but also constrains the effectiveness of entanglement-based metrological enhancements~\cite{ZhangQuantumEnhanced2023}. Parity doublets residing in the ground vibrational manifold would remove this ceiling entirely, as measurements become coherence-limited rather than lifetime-limited~\cite{Kozyryev2017PolyEDM,YuProbing2021,AugenbraunMolecular2020}. 

Asymmetric top molecules (ATMs) provide exactly this structure~\cite{AugenbraunMolecular2020}. In contrast to symmetric top molecules with two equal moments of inertia, ATMs are characterized by three distinct and independent moments of inertia. This combines with $3n-6$ vibrational degrees of freedom for $n$ constituent nuclei, resulting in a rich internal structure. When brought under control, the ATM structure can be exploited to provide laser cooling transitions, long-lived science states, large EDM sensitivity, and parity doublets for quantum control and systematic error rejection~\cite{AugenbraunMolecular2020}. In particular, ATMs generically support near-degenerate opposite-parity $K$-doublets already in the vibronic ground state, avoiding the finite lifetime of vibrationally excited $\ell$-doublet states and enabling science-state lifetimes expected to be larger than $ 10~\mathrm{s}$~\cite{Kozyryev2017PolyEDM,AugenbraunMolecular2020,YuProbing2021}. For example, in SrNH$_2$ the opposite-parity doublet in the $N=1$, $K_a=1$ manifold is split by $\sim 130~\mathrm{MHz}$, small enough to be polarized with modest electric fields while remaining long-lived~\cite{frenettVibrationalBranchingFractions2024,ThompsenRotational2000}. Laser cooling schemes for ATMs have been proposed~\cite{AugenbraunMolecular2020}, and laser cooling and photon cycling have recently been demonstrated in CaNH$_2$~\cite{li2026photoncyclinglasercooling}. Together with spectroscopy of vibrational branching ratios across a variety of ligands~\cite{Kozyryev2018}, these developments open the door to quantum control of ATMs.

Compared to other nonlinear geometries, the $C_{2v}$ geometry represents a middle ground between complexity and simplicity for laser cooling and quantum experiments. In comparison, molecules with only a single mirror plane (e.g., bent triatomics like $\mathcal{M}$--SH with $C_s$ symmetry) lack generic symmetry protection from vibrational and rotational mixing, resulting in more decay pathways and added experimental complexity~\cite{AugenbraunMolecular2020,augenbraunDirectLaserCooling2023,frenettVibrationalBranchingFractions2024}. Meanwhile, for symmetric top molecules (e.g.\ \mbox{$\mathcal{M}$--OCH$_3$} with $C_{3v}$ symmetry) with two equal moments of inertia, $\mathcal{N}$ reflection planes, and $\mathcal{N}$-fold axial rotational symmetry, vibrational branching is more manageable compared to $C_s$. However, these species exhibit doubly degenerate electronic states, which introduces sensitivity to vibronic perturbations. These perturbations can spontaneously break the symmetry of the top's two nominally identical moments of inertia (e.g., Jahn-Teller effects), complicating rotational branching for each vibrational decay path~\cite{Barckholtz01101998,Kozyryev2018}. This progression parallels the trade-off across optical-cycling ligands. Diatomic fluorides such as CaF and SrF cycle photons most simply and have reached advanced quantum control~\cite{anderegg_laser_2018,Shuman2010}, but offer no parity doublet; linear $\mathcal{M}$--OH species add a generic $\ell$-type parity doublet in the bending mode at the cost of vibrational structure~\cite{baum1DMagnetoOpticalTrap2020}, while the $C_{2v}$ metal amides provide a parity doublet in the vibronic \emph{ground} state, trading a modest increase in rotational complexity for the elimination of the vibrational-lifetime ceiling that limits $\ell$-doublet coherence~\cite{robichaud_parity-doublet_2026}.

In this paper, we investigate ATMs---specifically the $\mathcal{M}$--NH$_2$ family---for precision measurements of $P,T$-violating electromagnetic moments like electron EDM (eEDM). We present a comprehensive analysis that includes theoretical calculations and numerical simulations of eEDM sensitivity in ATMs. We extend the idea of engineered clock transitions~\cite{TakahashiEngineering2023,takahashi2025engineeredmolecularclocktransitions} to ATMs, showcasing how their unique structure allows for the engineering of specific rotational states that are highly sensitive to EDM effects while being robust against external perturbations~\cite{dereviankoColloquiumPhysicsOptical2011}. We further evaluate the differential ac Stark shifts induced by an optical dipole trap and identify the operating conditions under which trap-limited coherence supports the projected sensitivity.  Although we focus on the eEDM, the methodology developed here carries over directly to other $P,T$-violating observables; with a heavy-metal center such as radium, the same framework would enable searches for nuclear magnetic quadrupole and Schiff moments~\cite{gaulCPViolationSensitivity2024,Kudashovinitio2014}.

\section{Molecular Theory}

\begin{figure}[!htpb]
    \centering
    \includegraphics[width=0.5\linewidth]{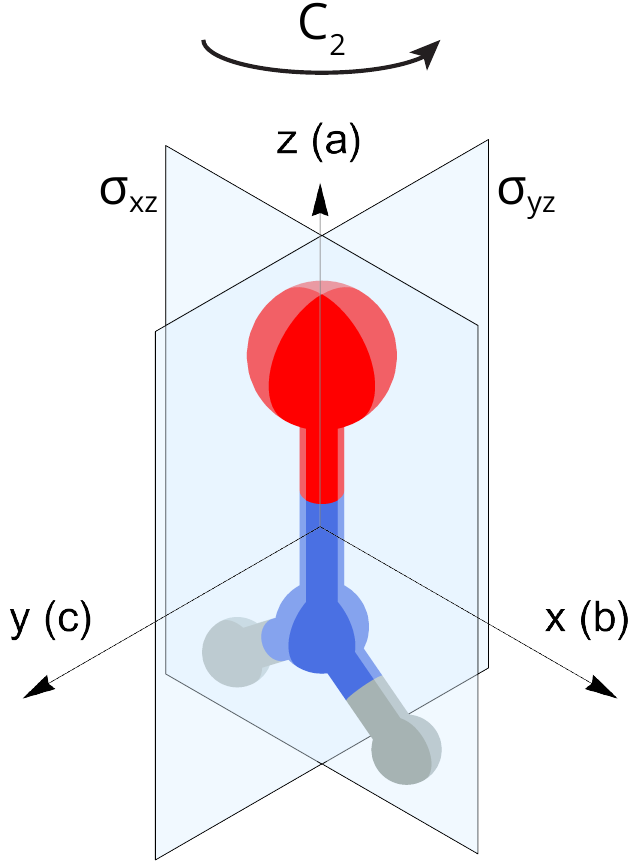}
    \caption{Diagram of the $C_{2v}$ point group structure of the molecules considered in this work. The red nucleus is a group 2 metal, blue is N, and gray is H. The axes label the molecular frame in terms of both the principal axes of rotation ($a,b,c$) and the quantization axes ($z,x,y$). The $C_{2v}$ operations are shown: $\pi$ rotation $C_2$, and reflection planes $\sigma_{xz}$ and $\sigma_{yz}$.}
    \label{fig:c2v-planes}
    \vspace{0.25cm} 
\end{figure}

\subsection{Symmetry and Overall Structure}

We consider ATMs of the form $\mathcal{M}$--NH$_2$ ($\mathcal{M}$ = Ca, Sr, Ba, Ra), which exhibit a planar equilibrium structure, invariant under $C_{2v}$ point group operations (see Figure~\ref{fig:c2v-planes}). Other metal centers such as Yb or Mg are expected to form analogous structures. The eigenstates of the molecule are classified according to their behavior under the molecular symmetry (MS) group~\cite{bunkerMolecularSymmetrySpectroscopy1998}, labeled $C_{2v}(\text{M})$, and the operations of this group commute with the total molecular Hamiltonian. The irreducible representations (irreps) of the MS group label the wavefunctions and are used to compute symmetry properties and selection rules. For $C_{2v}(\text{M})$, the irreps are all one-dimensional and non-degenerate, preventing mixing of different symmetry states and simplifying molecular structure. By contrast, symmetric top groups like $C_{3v}(\text{M})$ can exhibit multi-dimensional irreps, corresponding to degenerate representations not protected from state mixing. 

The action of the MS group can be separated into the actions of separate groups for rotational ($D_2$ point group), vibronic ($C_{2v}$ point group), and nuclear-spin degrees of freedom ($S_2$ permutation group). The action of these groups on the irreps of $C_{2v}(\text{M})$ is described by the characters listed in Table~\ref{tab:c2v-char} of Appendix~\ref{sec: group theory}. Further details of the symmetry structure of rotational and nuclear states are also included in Appendix~\ref{ap:MS_group}. 

\begin{figure*}[ht!]
    \centering
    \includegraphics[width=\textwidth]{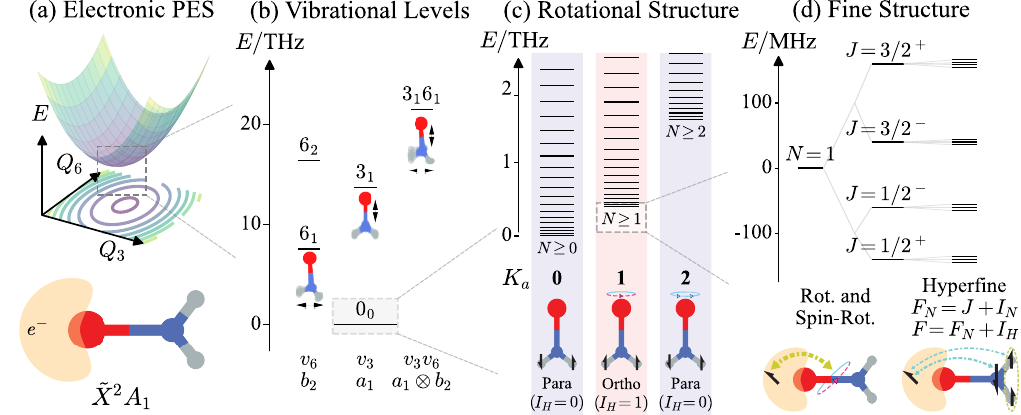}
    \caption{Molecular structure for $\mathcal{M}$--NH$_2$ asymmetric top free radicals discussed in this work. Within each sub-panel, energy splittings are approximately to scale for SrNH$_2$. (a) Electronic structure, showing a schematic potential energy surface (PES) for the electronic energy as a function of generalized vibrational coordinates $Q_6$ and $Q_3$, corresponding approximately to the NH$_2$ bending mode and $\mathcal{M}$--N stretching mode. Below the PES is a schematic of the valence $\tilde X {}^2 A_1$ electronic orbital, used for laser cooling and probing new physics. (b) A subset of the lowest-lying vibrational levels, labeled $n_{v_n}$ with $n=1,2\ldots 6$ denoting the vibrational mode, and $v_n$ the quanta in that mode. Here, $v_6$ denotes the molecular bending mode in the $ab$ plane with $b_2$ symmetry, $v_3$ denotes the $\mathcal{M}$--N stretching mode with $a_1$ symmetry, and the combined bending and stretching mode is denoted $v_3 v_6$ with symmetry $a_1 \otimes b_2$. (c) The rotational levels are grouped by different $|K|$ magnitudes, separated by $\gtrsim 0.1$~THz. For a given $|K|$, $N$ exhibits values $N \geq |K|$, with $N$ level splittings on $\gtrsim10$~GHz energy scales. Due to nuclear spin symmetry, the $K_a$ even (odd) states are paired with $I_H$ singlet (triplet) states, where $I_H = I_{H1}+I_{H2}$. (d) Further fine-structure splittings arise at $\lesssim 100$~MHz scales, due to spin-rotation interactions of the electron spin and molecular rotation, forming the quantum number $J$, parity doubling, which splits $\pm$ parities, and hyperfine interactions between the electron spin and nuclear spins $I_N$ and $I_H = I_{H1}+I_{H2}$,  adding up to total angular momentum $F$.}
    \label{fig:structure}
\end{figure*}

An overview of the relevant molecular structure of $\mathcal{M}$--NH$_2$ molecules is shown in Figure~\ref{fig:structure}. While the various molecular degrees of freedom span multiple orders of magnitude, the effective Hamiltonian approach allows us to focus on the ground vibronic states, denoted $\tilde X {}^2 A_1$. Here, the superscript indicates electron spin multiplicity, and the $A_1$ irrep labels the electronic and vibronic states, analogous to the ${}^2\Sigma^+$ label for linear molecules. We focus on the rotational and hyperfine structure most relevant for precision measurements and ignore other vibrational states, which are well separated from the vibronic ground state. Due to the small mass of the off-axis H atoms, the rotational structure is similar to that of a prolate symmetric top in Hund's case (b), with total rotational angular momentum $N$ and angular momentum projection $K=K_a$ along the $\mathcal{M}$--N bond, corresponding to the $a$-axis (see Figure~\ref{fig:c2v-planes}). 

Importantly, due to the identical pair of fermionic $I_{H,1}=I_{H,2}=1/2$ nuclear spins on the NH$_2$ ligand, the overall molecular wavefunction must be odd under H spin exchange. Since the ground vibronic state is a fully symmetric $A_1$ configuration, we focus our attention on how spin-symmetries correlate the rotational and nuclear symmetries. We describe the nuclear spins in the coupled basis with the total spin $\mathbf{I}_H = \mathbf{I}_{H1} + \mathbf{I}_{H2}$. The $I_H=0$ singlet is antisymmetric under exchange, and therefore must pair up with the even $K_a$ rotational states, which are symmetric under exchange. Meanwhile, the $I_H=1$ triplet states are exchange-symmetric and pair up with odd $K_a$ states. 

As a result, $K_a=1$ states are metastable, since decay to $K_a=0$ also requires the hydrogen spins to change from the triplet to the singlet. Only a small antisymmetric part of the hydrogen hyperfine interaction mixes these manifolds. In Appendix~\ref{sec:lifetime}, we estimate the lifetimes of the low-lying rotational levels in the $K_a$ = 1 and 2 stacks to be well over one year. Still, these intrinsic radiative lifetimes will be limited by thermal blackbody radiation, which is a concern for many trapped atom and molecule experiments. At 300~K, the lifetime from stimulated thermal processes is estimated at $\sim2$~sec, while at 77~K it increases to $\sim10^3$~s (see Appendix~\ref{sec:lifetime}). Therefore, the coherence times considered here with $\tau\gtrsim10$~s presume a cryogenic or shielded trap environment. 

\subsection{The Effective Hamiltonian}
\label{sec:eff_ham}

We use the effective Hamiltonian approach to describe the level structure of ATMs in a single vibronic state~\cite{Searscalculation1984}. We focus on the $\widetilde{X}^2A_1$ vibronic ground state, and the Hamiltonian includes interactions from rotation, spin-rotation, hyperfine, external field shifts, and the $P,T$-violating eEDM interaction (which can be amended to include other $P,T$-violating interactions as well). The Hamiltonian is given as a summation of each individual coupling sub-part. From top to bottom are rotation (R), (electron) spin-rotation (SR), hyperfine (HFS), Stark ($\mathcal{E}$), Zeeman ($\mathcal{B}$), and EDM interactions:
\begin{align}
        H_\text{R}        &= AN_a^2 + BN_b^2 +CN_c^2 \nonumber \\
        &= \sum_{k=0,2} {T^k(B_\text{rot})\cdot T^k(\mathbf{N},\mathbf{N}),}\\
        H_{\text{SR}}     &= \frac{1}{2}\sum_{\alpha=a,b,c} \epsilon_{\alpha\alpha} (N_\alpha S_\alpha + S_\alpha N_\alpha) \nonumber \\
                          &= \frac{1}{2}T^2(\bm{\epsilon})\cdot \left(T^2(\mathbf{N},\mathbf{S})+T^2(\mathbf{S},\mathbf{N})\right), \\
        H_{\text{HFS}}    &= \sum_{i} \left(a_F \mathbf{I}_i \cdot \mathbf{S} + \mathbf{I}_i \cdot \mathbf{T}_i \cdot \mathbf{S} \right) \nonumber \\
                          &= \sum_{i} \left(a_{F}^{(i)} T^1(\mathbf{I}_i) \cdot T^1(\mathbf{S})
                          + T^2(\mathbf{T}_i)\cdot  T^2(\mathbf{I}_i,\mathbf{S})\right),\\
        H_{\mathcal{E}}  &= -\mathbf{D} \cdot \bm{\mathcal{E}},\\
        H_{\mathcal{B}}  &= -\bm{\mu}\cdot\bm{\mathcal{B}},\\
        H_{\text{EDM}}   &= d_e W_d \, \mathbf{S} \cdot \hat{a} = d_e W_d \, T^1_{q=0}(\mathbf{S}).
\end{align}

Here $\mathbf{D}=D_0\hat{a}$ is the molecule-frame electric dipole moment, fixed along the $\hat{a}$ axis by symmetry, and $\bm{\mu} = -\mu_B\left(g_S\mathbf{1}+\tilde{\mathbf{g}}\right)\cdot\mathbf{S}+ \mu_N\left(g_H\mathbf{I}_H+g_N\mathbf{I}_N\right)$
is the total magnetic moment and $\bf \tilde g$ represents the anisotropic $g$-factor in the molecule frame. We ignore the contribution from further subleading effects at the ${\rm kHz}$ scale, such as the rotational g-factor contribution and the nuclear magnetic shielding tensor. The electronic sensitivity to the electron EDM is parameterized by $W_d$. More Hamiltonian details and matrix elements can be found in Appendices~\ref{ap:MS_group} and~\ref{ap:hamiltonian}, including our convention for defining $W_d$. 

For molecules containing $^{14}$N with nuclear spin $I_N=1$, an electric quadrupole term has to be added as well:
\begin{align}
    H_Q = \mathbf{I}_N \cdot \mathbf{Q} \cdot \mathbf{I}_N = eT^2(\mathbf{Q})\cdot T^2(\nabla \bm{\mathcal{E}}_\text{internal}),
\end{align}
where $T^2(\mathbf{Q})$ is related to the nuclear quadrupole moment and $T^2(\nabla \bm{\mathcal{E}}_\text{internal})$ are the components of the electric field gradient at the nucleus (see Appendix \ref{ap:hamiltonian} for matrix elements).

The electron spin--hydrogen hyperfine can be refactored in terms of the total coupled hydrogen spin, $\mathbf{I}_H = \mathbf{I}_{H,1} + \mathbf{I}_{H,2}$, and the difference spin, $\mathbf{I}_\Delta =  \mathbf{I}_{H,1} - \mathbf{I}_{H,2}$, resulting in: 
\begin{align}
    \begin{split}
        &\mathbf{I}_{H,1} \cdot \mathbf{T}_1 \cdot \mathbf{S} + \mathbf{I}_{H,2} \cdot \mathbf{T}_2 \cdot \mathbf{S}\\
    &= \mathbf{I}_H \cdot \frac{\mathbf{T}_1+\mathbf{T}_2}{2} \cdot \mathbf{S} 
    + \mathbf{I}_\Delta \cdot \frac{\mathbf{T}_1-\mathbf{T}_2}{2}\cdot \mathbf{S}.
    \end{split}
\end{align}
Here, $\mathbf{T}_1$ and $\mathbf{T}_2$ represent rank-2 hyperfine dipolar tensors for the two hydrogen nuclei. The first term preserves $I_H$ and therefore conserves nuclear spin symmetry, while the $I_\Delta$ term can mix $I_H$ and ortho/para states. 

Due to $C_{2v}$ symmetry, the Fermi contact constant $a_F$ and dipolar terms $T^2_{q=0,\pm 2}(\mathbf{T})$ are identical for the two hydrogen atoms, and the corresponding $I_\Delta$ term vanishes. The difference does not vanish for the dipolar $T_{ab} = T^2_{q=\pm1} (\mathbf{T})$ component, which transforms like $B_2$ and can mix $\Delta K_a =\pm 1$ states. Therefore, the $I_\Delta$ term preserves the overall fermion exchange symmetry by simultaneously changing both the symmetry of the nuclear and rotational wavefunctions. The $T_{ab}$ term has previously been determined in NF$_2$~\cite{MULLER2008185} and NH$_2$~\cite{Steimle1980Microwave}. It causes state mixing, suppressed in these near-prolate systems by $\sim \! T_{ab}/A \approx 3 \times 10^{-6}$, the ratio between the hyperfine coupling scale and the rotational separation of neighboring $K_a$ states. Due to this suppression, we do not include $T_{ab}$ in the effective Hamiltonian diagonalized below. 

Due to their lack of cylindrical symmetry, ATMs are typically described in the Hund's case (b) basis, with the electron spin quantized in the laboratory frame, and the spin-orbit interaction absorbed into the effective spin-rotation interaction. The rotational eigenstates are described using the symmetric top basis, $|N,K,M\ket$, which describe simultaneous eigenstates of the operators $\mathbf{N}^2$, $\mathbf{N}\cdot \hat{a}$, and $\mathbf{N}\cdot \hat{Z}$, with eigenvalues $N(N+1)$, $K$, and $M$ respectively. The projection quantum numbers $K$ and $M$ are defined in terms of quantization axes in the molecular ($\hat a$) and laboratory ($\hat Z$) frames, respectively. 

For unbroken cylindrical symmetry, the states $|N,\pm K,M\ket$ form a degenerate parity doublet. In $C_{2v}$ systems, the degeneracy is lifted by the rotational asymmetry term, giving rise to the parity eigenstates in the absence of external fields,
\begin{align}
    |N\!,\!M\!,\!\pm  \ket = \frac{1}{\sqrt{2}}(|N\!,\!K\!,\!M \ket \pm (-1)^{N-K} |N\!,\!-K\!,\! M \ket).
\end{align}
The fine and hyperfine interactions also generically break cylindrical symmetry and further split the parity doublet. Fine structure is described by $\mathbf{J} = \mathbf{N}+\mathbf{S}$, where $\mathbf{S}$ stands for valence electronic spin, and hyperfine by $\mathbf{F} = \mathbf{J}+\sum_i\mathbf{I}_i$, where $F$ is the total angular momentum quantum number, and $\mathbf{I}_i$ is each individual nuclear spin. 

The application of an external static magnetic field couples the various molecular angular momenta to the laboratory frame, splitting the degeneracy of Zeeman sublevels. For an electron in an unperturbed $s\sigma$-like $a_1$ molecular orbital, the spin is decoupled from the molecular frame, and the magnetic spin Zeeman interaction is isotropic, similar to that of a free electron. However, due to cross-perturbations with orbital Zeeman and spin-orbit interactions, the electron spin can couple to the molecule's internal angular momenta, and then the effective $g$-factor becomes anisotropic in the molecular frame. Similar cross-perturbations of spin-orbit and Coriolis interactions are also responsible for the effective spin-rotation coupling. This similarity provides an approximate relationship between the electron spin-rotation constant $\epsilon$ and the anisotropic magnetic $g$-factor, first derived by Curl~\cite{Curlrelationship1965}. The effective Hamiltonian for the Zeeman interaction in doublet electronic molecular states is detailed in Refs.~\cite{brownRotationalSpectroscopyDiatomic2003,Searscalculation1984} and, assuming negligible kinetic spin Zeeman (SZ-K) and external-magnetic-field-induced spin–orbit (B-SO) contributions\cite{melo2003}, we have for the $g$-factor anisotropy 
\begin{align}\label{eq:curl}
\tilde g_{\alpha\alpha} =g_{\alpha\alpha}-g_S
\hspace{2mm}\simeq\hspace{2mm}
\tilde g^\text{Curl}_{\alpha\alpha} = - \frac{\epsilon_{\alpha\alpha}}{2B_{\alpha\alpha}} \,,
\end{align}
where $\alpha$ denotes an axis of the molecular frame, $\tilde g$ is anisotropy that remains after subtracting the free-electron spin $g$-factor. 

To demonstrate an EDM measurement scheme in ATMs, we simulate the rotational level structure in the ground electronic and vibrational manifold of strontium and radium monoamides (Sr/Ra$^{14}$NH$_2$ and Sr/Ra$^{15}$NH$_2$). Strontium is chosen as an example since most of the $\mathcal{M}$--NH$_2$ species have very similar structure apart from relatively minor differences in molecular constants. Radium, however, represents a standout example because of its fundamentally different spin-rotation structure due to its extraordinarily large spin-orbit constant. The states are represented in the Hund's case (b) basis, defined as $|N,K;J(N,S),F_N(J,I_N),F(F_N,I_H), M \ket$, where $I_N$ is the nitrogen spin and $I_H=I_{H,1} + I_{H,2}$ is the total spin of the coupled fermionic hydrogen system. $^{14}$N is the most naturally abundant isotope (99.6\% abundance) with $I_N=1$, which results in fermionic total angular momentum states with half-integer $F$ and adds an additional nuclear quadrupole hyperfine structure. We also consider $^{15}$N (0.4\% abundance), with $I_N=1/2$, resulting in bosonic total angular momentum states with integer $F$ and simpler hyperfine structure with only dipolar interactions.

The rotational and spin-rotational structure of SrNH$_2$ in the ground electronic state, $\widetilde{X}{}^2 A_1$, has previously been extensively investigated in Ref.~\cite{ThompsenRotational2000}, providing a solid foundation for accurately modeling the molecular structure. However, the hyperfine structure remained unresolved in these high-temperature studies. Here, we have performed theoretical calculations to obtain predicted parameters for the hyperfine effective Hamiltonian. Table~\ref{tab:constants} includes effective Hamiltonian parameters from both experiment and theory for a variety of alkaline-earth monoamide molecules. The parameters include an anisotropic dipolar tensor term $T_{ab}$ for individual hydrogen parts valued $\approx 1$~MHz (except for BaNH$_2$, where $T_{ab} \approx 0.06$~MHz). In particular, the (2, 1) tensor component mixes rotational levels with $\Delta K$ = 1, thereby mixing the ortho and para states. However, given that it is an order of magnitude weaker than spin-rotation mixing, this does not affect the general utility of the science state, and our simulation supports this claim.

\section{Molecular Constants from Relativistic ab-initio Calculations}
\label{sec:ab-initio-calc}

\begin{table*}[t]
    \centering
    \caption{Ground state $\widetilde{X}^2\text{A}_1$ molecular constants (MHz), shifted electronic $\tilde g$-factors ($10^{-3}$), and calculated molecular-frame electric dipole moments ($D_0$). Here $^{14}$N, $^{15}$N, and $^{1}$H denote isotope-specific hyperfine parameters for nitrogen and hydrogen. Unless stated otherwise, the given values are obtained from \textit{ab-initio} calculations and correspond to the $\mathcal{M}$--$^{14}$NH$_2$ isotopologue. Experimental rotation and spin-rotation parameters of CaNH$_2$ and SrNH$_2$~\cite{Brewsterpure2000,SheridanRotational2005a} are given for comparison and supersede calculated parameters in further modeling.
    \textit{Ab-initio} relativistic values of $\tilde g_{\alpha\alpha}$ are compared with approximate values $\tilde g^\text{Curl}_{\alpha\alpha}$ obtained from Curl relation Eq.~\eqref{eq:curl} using theoretical and experimental values of the spin-rotation constants.}
    \label{tab:constants}
    \begin{ruledtabular}
    \begin{tabular}{lcccc}
        Parameter & CaNH$_2$ & SrNH$_2$ & BaNH$_2$ & RaNH$_2$ \\
        \hline
        $A$\hspace{4mm}exp.    & 392127\phantom{.000} & 394353\phantom{.000} &  &   \\
           \hspace{6.5mm}calc. & 391566\phantom{.000} & 394340\phantom{.000} & 385251\phantom{.000} & 389324\phantom{.000}  \\[0.5ex]
        \graymid
        $B$\hspace{4mm}exp.    & 9009.1               & 6790.3 &    &     \\
           \hspace{6.5mm}calc. & 9005.7               & 6780.1 & 5783.7   & 5328.3    \\[0.5ex]
        \graymid
        $C$\hspace{4mm}exp.    & 8782.8               & 6659.5 &  &    \\
           \hspace{6.5mm}calc. & 8803.2               & 6665.5 & 5698.2   & 5256.4    \\[1ex]
        \graymid
        
        $\epsilon_{aa}$\hspace{2.3mm}exp. & \phantom{00}45.7 & \phantom{0}160.1 &                  &  \\
        \hspace{6.5mm}calc.               & \phantom{00}25.2 & \phantom{00}80.4 & \phantom{0}195.2 & \phantom{0}496.0 \\[0.5ex]
        \graymid
        $\epsilon_{bb}$\hspace{2.8mm}exp. & \phantom{00}32.1 & \phantom{00}59.7 &                  &  \\
        \hspace{6.5mm}calc.               & \phantom{00}37.2 & \phantom{00}67.4 & \phantom{00}52.6 & \phantom{0}104.1 \\[0.5ex]
        \graymid
        $\epsilon_{cc}$\hspace{2.8mm}exp. & \phantom{00}41.1 & \phantom{00}89.7 &                  &  \\        
        \hspace{6.5mm}calc.               & \phantom{00}48.5 & \phantom{0}105.3 & \phantom{0}119.3 & \phantom{0}278.6 \\[1ex]      
        \graymid
        
        $a_F\;(^{14}\text{N}\,|\,^{15}\text{N}\,|\,^1\text{H})$
          & \phantom{--}21.1 $|$ --30.0 $|$ \phantom{--}2.8 
          & \phantom{--}18.4 $|$ --25.8 $|$ \phantom{--}2.2
          & \phantom{0--}9.6 $|$ --13.4 $|$ \phantom{--}2.2  
          & \phantom{--}15.8 $|$ --22.1 $|$ \phantom{--}2.3 \\
          \graymid
        $T_{aa}(^{14}\text{N}\,|\,^{15}\text{N}\,|\,^1\text{H})$
          & \phantom{0--}1.5 $|$ \phantom{0}--2.1 $|$ \phantom{--}4.2 
          & \phantom{0--}1.2 $|$ \phantom{0}--1.7 $|$ \phantom{--}3.4
          & \phantom{0--}0.5 $|$ \phantom{0}--0.8 $|$ \phantom{--}0.5 
          & \phantom{0--}1.0 $|$ \phantom{0}--1.5 $|$ \phantom{--}1.8 \\
          \graymid
        $T_{bb}\,(^{14}\text{N}\,|\,^{15}\text{N}\,|\,^1\text{H})$
          & \phantom{0}--0.5 $|$ \phantom{--0}0.7 $|$ --2.1 
          & \phantom{0}--0.4 $|$ \phantom{--0}0.5 $|$ --2.0
          & \phantom{0--}0.0 $|$ \phantom{0}--0.1 $|$ --0.1 
          & \phantom{0}--0.3 $|$ \phantom{--0}0.4 $|$ --0.8 \\
          \graymid
        $T_{cc}\,(^{14}\text{N}\,|\,^{15}\text{N}\,|\,^1\text{H})$
          & \phantom{0}--1.0 $|$ \phantom{0--}1.4 $|$ --2.2 
          & \phantom{0}--0.9 $|$ \phantom{0--}1.2 $|$ --1.3
          & \phantom{0}--0.6 $|$ \phantom{0--}0.8 $|$ --0.4 
          & \phantom{0}--0.8 $|$ \phantom{0--}1.1 $|$ --1.0 \\
          \graymid
        $T_{ab}\,$($^{1}$H) & \phantom{0.}$\pm$1.5 & \phantom{0.}$\pm$1.1 & \phantom{0.--}$\pm$0.06 & \phantom{0.}$\pm$1.0 \\[1ex]     \graymid   
          
        $\chi_{aa}$($^{14}$N) & \phantom{--00}3.3 & \phantom{--00}3.3 & \phantom{--00}3.2 & \phantom{--00}3.2 \\ 
        \graymid
        $\chi_{bb}\,$($^{14}$N) & \phantom{00}--3.1 & \phantom{00}--3.0 & \phantom{00}--3.0 & \phantom{00}--3.1 \\
        \graymid
        $\chi_{cc}\,$($^{14}$N) & \phantom{00}--0.2 & \phantom{00}--0.3 & \phantom{00}--0.2 & \phantom{00}--0.2 \\[1ex]
        \graymid
        
        $\tilde g_{aa}$ & \phantom{000}--0.09& \phantom{000}--0.31& \phantom{000}--0.75&\phantom{000}--4.49 \\
        $\tilde g^\text{Curl}_{aa}$\hspace{2mm}calc. & \phantom{000}--0.03& \phantom{000}--0.10& \phantom{000}--0.25&\phantom{000}--0.64\\
        \hspace{9mm}exp. & \phantom{000}--0.06& \phantom{000}--0.20& &\\ [0.5ex]
        \graymid
        $\tilde g_{bb}$ & \phantom{000}--2.11& \phantom{000}--5.08& \phantom{000}--4.71&\phantom{00}--10.48\\
        $\tilde g^\text{Curl}_{bb}$\hspace{2mm}calc. & \phantom{000}--2.07& \phantom{000}--4.97& \phantom{000}--4.55&\phantom{000}--9.77\\
        \hspace{9mm}exp. & \phantom{000}--1.78& \phantom{000}--4.40& &\\ [0.5ex]
        \graymid
        $\tilde g_{cc}$ & \phantom{000}--2.80& \phantom{000}--8.01& \phantom{00}--10.62&\phantom{00}--27.25\\
        $\tilde g^\text{Curl}_{cc}$\hspace{2mm}calc. & \phantom{000}--2.75& \phantom{000}--7.90& \phantom{00}--10.47&\phantom{00}--26.50\\
        \hspace{9mm}exp. & \phantom{000}--2.34& \phantom{000}--6.73& &\\[1ex]
        \graymid

                $D_0$(Debye) & \phantom{--00}1.37 & \phantom{--00}1.86 & \phantom{--00}4.41 & \phantom{--00}4.19 \\
    \end{tabular}
    \end{ruledtabular}
\end{table*}

The magnetic hyperfine tensor elements, electronic $g$-factors, electron spin-rotation couplings, and molecular-frame electric dipole moments were calculated using the relativistic Fock-space coupled-cluster method with single and double excitations (FS-CCSD)~\cite{visscherFormulationImplementationRelativistic2001} in the sector (0,1). For the magnetic hyperfine tensors, $g$-factors, and spin-rotation couplings, the dyall.cv3z basis set was employed~\cite{dyallRelativisticDoublezetaTriplezeta2016,dyallRelativisticDoubleZetaTripleZeta2009}, together with a large active space spanning orbital energies between $\pm800$~a.u.
The molecular-frame electric dipole moments were computed under the same computational conditions, except that the dyall.av3z basis set was used instead.

In addition, the nuclear quadrupole couplings for $^{14}$N were calculated using density functional theory (DFT) with the dyall.cv4z basis set~\cite{dyallRelativisticDoublezetaTriplezeta2016,dyallRelativisticDoubleZetaTripleZeta2009} and the CAM-B3LYP* functional, which is optimized for calculations of electric field gradients~\cite{thierfelderCu63andAu197nuclearQuadrupoleMoments2007}.

All calculations were performed in a four-component Dirac--Coulomb relativistic framework within the electronic structure program package DIRAC~\cite{saueDIRAC25DIRACRelativistic2025}, using Gaussian-type finite nuclear models to describe the electron-nucleus Coulomb interaction. For the calculations of the electronic $g$ factors, the gauge origin of the magnetic vector potential associated with the external magnetic field was chosen at the molecular center of mass, such that the vector potential vanishes at that point.

Given the discrepancies observed in Table~\ref{tab:constants} between experimental and \textit{ab-initio} theoretical values of $\epsilon_{aa}$ for CaNH$_2$ and SrNH$_2$, we carried out a more detailed investigation of these systems by means of four-component relativistic single-reference coupled-cluster calculations including single, double, and perturbative triple excitations, CCSD(T), using the dyall.cv4z basis set and an active space spanning orbital energies between $\pm800$~a.u.
The sensitivity of $\epsilon_{aa}$ to the theoretical description is much higher compared to $\epsilon_{bb}$ and $\epsilon_{cc}$ due to the very high rotational constants $A$.
The obtained values are:
$\epsilon_{aa} =  55.9$~MHz,
$\epsilon_{bb} =  36.2$~MHz, and
$\epsilon_{cc} =  46.7$~MHz for CaNH$_2$, and
$\epsilon_{aa} = 118.1$~MHz,
$\epsilon_{bb} =  66.8$~MHz, and
$\epsilon_{cc} = 100.9$~MHz for SrNH$_2$.
These results indicate that electron correlation effects play a critical role in the description of this parameter. This conclusion is further supported by the observed progression from CCSD to CCSD(T), which systematically shifts the calculated values toward the experimental results. Owing to the high computational cost associated with this level of theory for systems containing heavy elements such as Ba and Ra, it was intractable to use this method for BaNH$_2$ and RaNH$_2$. Therefore, we have chosen to report a consistent set of values for all four systems obtained using the FSCC method (as described above) in Table~\ref{tab:constants}.

Geometries for CaNH$_2$ and SrNH$_2$ used in the calculations were sourced from available experimental data~\cite{Brewsterpure2000,SheridanRotational2005a,ThompsenRotational2000}. 
We note here that the CaNH$_2$ and SrNH$_2$ geometries derived in the original works~\cite{Brewsterpure2000,SheridanRotational2005a,ThompsenRotational2000} showed large deviations from the measured rotational constants, prompting a refit of the structure, as discussed in more detail in Appendix~\ref{ap:refit}.
For RaNH$_2$, the available high-level theoretical geometry was used~\cite{zhangAnalyticGradientsRelativistic2023} (note the misprint in that Reference, mentioned in Appendix~\ref{ap:refit}). Finally, the BaNH$_2$ geometry was optimized using quasi-relativistic DFT using the B3LYP functional~\cite{stephensInitioCalculationVibrational1994} and the Def2-TZVPP basis set~\cite{kauppPseudopotentialApproachesCa1991,weigendBalancedBasisSets2005}, which was able to reliably reproduce the geometries of the other three molecules to a high degree of accuracy -- errors in bond lengths and angles at the level of 0.01~\AA\ and 1\degree, respectively (Table~\ref{tab:geom} in Appendix~\ref{ap:refit}).

Using the four-component relativistic formalism described above, we calculated the $g_{\alpha \alpha}$ values and, therefore, also $\tilde g_{\alpha\alpha}$ (see Eq.~\eqref{eq:curl}). In addition, following the theoretical formalism outlined in Ref.~\cite{aucarTheoreticalStudyNuclear2012}, we computed the relativistic values of the electron spin-rotation tensor elements. Combined with the Curl relationship given in Eq.~\eqref{eq:curl}, which is based on a semi-relativistic treatment including only the leading-order spin-orbit relativistic contribution in a $1/c^2$ expansion, these results yielded the corresponding $\tilde g^\text{Curl}$ values. Furthermore, available experimental spin-rotation constants (Table~\ref{tab:constants}) were also used to derive experimental $\tilde g^\text{Curl}$ values.
The small ratios $|\tilde g^\text{Curl}_{\alpha\alpha}|=|\epsilon_{\alpha\alpha}|/(2B_{\alpha\alpha}) \ll 1$
indicate weak spin--rotation coupling, supporting the approximation that these molecules behave magnetically like an effective single-electron spin, with modest molecule- and state-dependent corrections.

While \textit{ab-initio} calculations are needed to obtain quantitatively accurate results for the $g$-factors of the $C_{2v}$ molecules we consider here, it can be seen in Table \ref{tab:constants} that the approximate values obtained using the Curl relationship can capture the competition between magnetic couplings in the laboratory frame and $g$-factor anisotropy in the molecular frame. The largest discrepancy between $\tilde g_{aa}$ and $\tilde g^\text{Curl}_{aa}$ values can be observed in the case of RaNH$_2$, highlighting the importance of the higher-order relativistic contributions in the case of heavy, EDM-sensitive species.

\section{Engineered Clock Transitions for EDM Measurements}

\subsection{Electromagnetic Field Sensitivities}

Recently, Refs. \cite{AndereggQuantum2023, TakahashiEngineering2023,takahashi2025engineeredmolecularclocktransitions} showed that there exist optimal electric field values that produce states and transitions with significant $CP$-violation sensitivity while being insensitive to external field noise. Instead of performing EDM measurements with spin precession of stretched states in the fully polarized molecule regime, this method uses external fields to mix states and engineer noise-insensitive clock transitions with EDM sensitivity. 

To identify and characterize these transitions in $\mathcal{M}$--NH$_2$ molecules, we search for pairs of states with zero-crossings of the differential first-order field sensitivity that retain finite EDM sensitivity. We investigate Sr$^{14,15}$NH$_2$ and Ra$^{14,15}$NH$_2$ in the $N=1$, $K_a=1$ ortho manifold, as it hosts the lowest-lying parity doublet, can be polarized at $\sim10^2$~V/cm fields, and is metastable due to nuclear spin-statistics. Doublets with larger $K_a$ have smaller splittings, while increasing $N$ has the effect of widening the doublet, providing tunability based on the states used.

The effective Hamiltonian describing the molecular structure is constructed from standard matrix elements in the literature (see the Appendix~\ref{ap:hamiltonian} and Ref.~\cite{Searscalculation1984}), and includes relevant interactions with external electric and magnetic fields $(\mathcal{E},\mathcal{B})$ directed along the lab $Z$-frame. 
The Hamiltonian can be written as a function of the applied fields, $H(\mathcal{E},\mathcal{B})$, and diagonalization provides parametric energy eigenvalues $E_i (\mathcal{E},\mathcal{B})$ and eigenvectors $|v_i(\mathcal{E},\mathcal{B})\rangle$, labeling the $i$-th eigenstate. At each $(\mathcal{E},\mathcal{B})$ point, we compute properties of interest that describe the states' sensitivity to external noise and to the EDM signal. 

For first-order properties, we use the Hellmann-Feynman theorem: $\partial_\lambda E_i=\langle\partial_\lambda H\rangle_i$ for the parameters $\lambda=\mathcal{E},\mathcal{B}$. Each derivative is the lab-frame projection of a molecular moment: $\partial_\mathcal{E} E = -\langle \mathbf{D}\cdot\hat{Z}\rangle$, and $\partial_\mathcal{B} E = -\langle \bm{\mu}\cdot\hat{Z}\rangle$.

Taking the difference between the two states of the transition, $i$ and $j$, gives the first-order sensitivities of the transition frequency $f= E_i-E_j$, given by $f_\lambda \equiv \partial_\lambda f = \langle\partial_\lambda H\rangle_i-\langle\partial_\lambda H\rangle_j$.
The EDM sensitivity of a state is the molecule-frame projection of the spin, $\Sigma=\mathbf{S}\cdot\hat{a}$, and the differential sensitivity of the transition is $\Delta\Sigma=\langle\Sigma\rangle_i-\langle\Sigma\rangle_j$. The transitions we seek are the points at which $f_\mathcal{E}=f_\mathcal{B}=0$ while $|\Delta\Sigma|$ remains large.

As $\mathcal{E}$ is increased to polarize the molecule, the orientation $\langle\mathbf{D}\!\cdot\!\hat Z\rangle$ and spin projection $\langle\Sigma\rangle$ asymptotically approach $1/2$ of their body-frame maxima, reflecting the sharing of rotational angular momentum between the molecular and laboratory frames~\cite{PetrovSensitivity2022,AugenbraunMethods2021}. 

When the first-order sensitivities are tuned to zero, second-order shifts dominate the residual noise sensitivity. 
To estimate these effects at second-order, we calculate the energy shift from mixing with nearby states induced by variation of parameters $\lambda$ and $\lambda^\prime$:
\begin{align}
    \partial_{\lambda\lambda^\prime}E_i = 2 \sum_{k\neq i} \frac{\langle i|\partial_\lambda H|k\rangle\langle k|\partial_{\lambda^\prime} H|i\rangle}{E_i-E_k}, \label{eq:second-order}
\end{align} 
with corresponding differential frequency curvature $f_{\lambda\lambda^\prime} \equiv \partial_{\lambda\lambda^\prime}E_i -\partial_{\lambda\lambda^\prime}E_j $ used to characterize the transition properties. This allows us to independently calculate both the slope and curvature of each transition at each point, bypassing issues arising from finite-difference grid size and differences in eigenstate labeling between neighboring points. More details on eigenstate calculations are provided in the Appendix~\ref{ap:tracking}.

\subsection{Optimal EDM Transitions}

\begin{table*}[t]
    \centering
    \renewcommand{\arraystretch}{1.15}
    \caption{Optimal EDM clock transitions in the $N=1, K_a=1$ manifold. For each species, we show both the finite $\mathcal{B}$ transition and time-reverse pair (states labeled with $^\ast$) that maximize the figure-of-merit $\mathcal{F}=|\Delta\Sigma|\sqrt{\tau_\text{EM}/1~{\rm s}}$, with an assumed field fluctuation of $\delta \mathcal{E} = 10~\text{mV/cm}$ and $\delta \mathcal{B} = 100$~\textmu G, divided equally between the longitudinal ($Z$) and transverse ($X$) directions ($\delta\lambda_{\text{axis}} = \delta\lambda_\text{total}/\sqrt{2}$). Here $\tau_{\mathrm{EM}}$ is the electromagnetic coherence time, $\Sigma$ the eEDM sensitivity, and $f$ the transition frequency. The last column gives the dominant noise term from Eq.~\ref{eq:deltaf}, with units $f_{\mathcal{E}}$ in MHz/(V/cm), $f_{\mathcal{B}}$ in MHz/G, $f_{\mathcal{EE}}$ and $f_{\mathcal{E}_\perp}$ in MHz/(V/cm)$^2$, and $f_{\mathcal{EB}}$ in MHz/(V/cm$\cdot$G), with $f_{\lambda_\perp}\equiv \operatorname{Var}_{\lambda_\perp}\!\left[f\right]$ (see Appendix~\ref{ap:tauem}). The $\Delta M=1$ time-reverse pairs ($\mathcal{M}{}^{14}$NH$_2$) are listed with a finite $\mathcal{B}$-field bias to reduce sensitivity to transverse noise, while the $\Delta M=2$ time-reverse pairs ($\mathcal{M}{}^{15}$NH$_2$) are protected to second-order and operate at much smaller bias field. For all transitions, $|\Delta\Sigma|$ is obtained by taking the magnitude of $\Delta \Sigma$ after averaging over the field noise at the operating point.}
    \label{tab:magic}
    \begin{ruledtabular}
    \begin{tabular}{llccccccl}
        Species & Transition States & $\mathcal{E}$ & $\mathcal{B}$ & $\tau_{\mathrm{EM}}$ &
        $|\Delta\Sigma|$ & $\mathcal{F}$ & $f$ & Limiting $\delta f$ \\
        & & (V/cm) & (G) & (s) & &  & (MHz) & \\
        \hline
        $^{88}$Sr$^{14}$NH$_2$ & $M(-\tfrac{1}{2}\!\to\!+\tfrac{1}{2})^*$ & 128.95 & 0.0104 & 2.0 & 0.17 & 0.25 & $4.8{\times}10^{-5}$ & $f_{\mathcal{B}}\!=\!4.6{\times}10^{-3}$\\
        $^{88}$Sr$^{14}$NH$_2$ & $M(+\tfrac{1}{2}\!\to\!+\tfrac{3}{2})$ & 145.35 & 28.26 & 65.1 & 0.57 & 4.63 & 55.0 & $f_{\mathcal{EE}}\!=\!-2.8{\times}10^{-4}$\\[3pt]
        $^{226}$Ra$^{14}$NH$_2$ & $M(-\tfrac{1}{2}\!\to\!+\tfrac{1}{2})^*$ & 193.64 & 0.0084 & 1.6 & 0.82 & 1.02 & $4.2{\times}10^{-5}$ & $f_{\mathcal{E}}\!=\!5.5{\times}10^{-5}$\\
        $^{226}$Ra$^{14}$NH$_2$ & $M(-\tfrac{3}{2}\!\to\!+\tfrac{1}{2})$ & 194.52 & 0.47 & 261.7 & 0.82 & 13.32 & 0.3 & $f_{\mathcal{EB}}\!=\!6.6{\times}10^{-3}$\\[3pt]
        $^{88}$Sr$^{15}$NH$_2$ & $M(-1\!\to\!+1)^*$ & 142.31 & $3.3{\times}10^{-5}$ & 711.0 & 0.13 & 3.48 & $2.5{\times}10^{-10}$ & $f_{\mathcal{EB}}\!=\!2.4{\times}10^{-3}$\\
        $^{88}$Sr$^{15}$NH$_2$ & $M(0\!\to\!+2)$ & 145.02 & 26.69 & 54.3 & 0.59 & 4.37 & 56.8 & $f_{\mathcal{E}\perp}\!=\!4.9{\times}10^{-4}$\\[3pt]
        $^{226}$Ra$^{15}$NH$_2$ & $M(-1\!\to\!+1)^*$ & 176.37 & $4.6{\times}10^{-5}$ & 232.1 & 0.72 & 10.93 & $1.3{\times}10^{-9}$ & $f_{\mathcal{EB}}\!=\!6.3{\times}10^{-3}$\\
        $^{226}$Ra$^{15}$NH$_2$ & $M(+1\!\to\!+2)$ & 179.53 & 1.57 & 343.8 & 0.49 & 9.10 & 1.6 & $f_{\mathcal{EB}}\!=\!4.2{\times}10^{-3}$\\
    \end{tabular}
    \end{ruledtabular}
\end{table*}

\begin{figure*}[htbp]   
\centering
\includegraphics[width=1\linewidth]{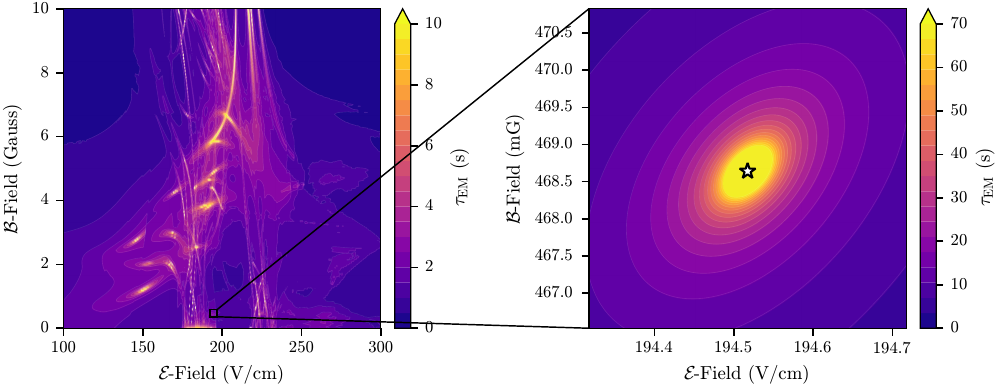} 
\caption{2-D map of engineered clock transitions, shown for $^{226}$Ra$^{14}$NH$_2$ in the $K_a = 1$ ortho manifold.  The color scale is the electromagnetic coherence time $\tau_{\mathrm{EM}}$ evaluated for quasistatic $\delta \mathcal{E}$ and $\delta \mathcal{B}$ field noise at the amplitudes of Table~\ref{tab:magic}. The left panel sums $\tau_{\mathrm{EM}}$ over the clock transitions with figure of merit $\mathcal{F} \geq 0.25$ and EDM sensitivity $|\Delta\Sigma| > 0.1$. Each bright region corresponds to $f_\mathcal{E}\approx f_\mathcal{B}\approx 0$. The right panel isolates the featured crossing $M(-\tfrac{3}{2} \to +\tfrac{1}{2})$ at $(194.52~\mathrm{V/cm},\, 0.47~\mathrm{G})$, listed in row 4 of Table~\ref{tab:magic}, with a star at the operating point that maximizes $|\Delta\Sigma|\sqrt{\tau_{\mathrm{EM}}}$, where $|\Delta\Sigma| = 0.82$ and $f = 0.3$~MHz. Across this zoom window, $|\Delta\Sigma|=0.82$ varies by less than $10^{-4}$. The residual field sensitivity of this crossing is dominated by the mixed derivative $f_{\mathcal{E}\mathcal{B}}$.  The zoom box on the left
panel is drawn enlarged and is not to scale.  Including the transverse response, as in Table~\ref{tab:magic}, lowers $\tau_{\mathrm{EM}}$ at this point from $284$~s to $262$~s.}
\label{fig:2D-magic}
\end{figure*}

To optimize transitions for eEDM measurements, we define the dimensionless figure of merit 
\begin{equation}
\mathcal F \equiv
|\Delta\Sigma|
\sqrt{\frac{\tau_{\rm EM}}{1\,\mathrm{s}}}.
\label{eq:fom}
\end{equation}
Here, $\tau_\text{EM}$ is the electromagnetic coherence timescale, set by external-field noise. With the variance of the transition frequency given by $\sigma_f^2\equiv\operatorname{Var}(\delta f)$, we define $\tau_{\rm EM}\equiv\sigma_f^{-1}$ (see Appendix~\ref{ap:tauem} for details). The $\sqrt{\tau_{\rm EM}}$ dependence results from assuming a coherence time-limited experimental repetition rate. 

For a specified noise budget, $\mathcal F$ isolates the
transition-specific contribution to the eEDM sensitivity. Assuming a coherence time $\tau=\tau_{\rm EM}$, the quantum-projection-noise-limited sensitivity is
\begin{equation}
    \delta d_e=\frac{\hbar}{W_d}\times\frac{1}{\mathcal{F}[{\rm s}^{1/2}]}\times \frac{1}{\sqrt{N_{\rm p}\mathcal{D} T_{\rm int}}},
\end{equation}
where $N_{\rm p}$ is the number of detected molecules per experimental pulse, $R$ is the pulse repetition rate,
$\mathcal D\equiv R\tau$ is the duty-cycle/multiplexing factor, and $T_{\rm int}$ is the total integration time. This expression separates the electronic enhancement $W_d$ and transition quality $\mathcal F$ from the apparatus throughput $N_{\rm p}\mathcal D$ and integration time. The quantity $N_{\rm p}\mathcal D T_{\rm int}=N_{\rm tot}\tau$ can be interpreted as the total detected molecule-interrogation time.

We focus on optimizing $\mathcal{F}$. We consider the first- and second-order contributions to frequency noise, $\delta f$, from independent, quasi-static fluctuations along parallel and transverse fields, $\delta \mathcal{E}_\parallel$, $\delta \mathcal{B}_\parallel$, $\delta \mathcal{E}_\perp$, $\delta \mathcal{B}_\perp$:
\begin{align}
    \operatorname{Var}\left[\delta f\right] &=
    ( f_{\mathcal{E}} \, \delta \mathcal{E}_\parallel)^2
    + ( f_{\mathcal{B}} \, \delta \mathcal{B}_\parallel)^2
    + ( f_{\mathcal{E}\mathcal{B}}\,\delta \mathcal{E}_\parallel\,\delta \mathcal{B}_\parallel)^2
    \nonumber \\ \nonumber
    &+ \tfrac{1}{2}( f_{\mathcal{E}\mathcal{E}}\,\delta \mathcal{E}_\parallel^2)^2
    + \tfrac{1}{2}( f_{\mathcal{B}\mathcal{B}}\,\delta \mathcal{B}_\parallel^2)^2
    \\ \label{eq:deltaf}
    & + \operatorname{Var}_{\delta \mathcal{E}_\perp}\!\left[f\right] 
    + \operatorname{Var}_{\delta \mathcal{B}_\perp}\!\left[f\right] ,
\end{align}
where $\operatorname{Var}_\lambda\!\left[\cdot\right]$ denotes the variance of the computed quantity over a normally distributed parameter $\lambda$. Numerical evaluation of the variance avoids divergences in Eq.~\ref{eq:second-order} encountered when summing over time-reverse pairs of states that become degenerate as $\mathcal{B}\rightarrow0$. This is primarily a concern for $\Delta M=1$ pairs, which are linearly sensitive to $\delta \mathcal{B}_\perp$ noise at $\mathcal{B}_\parallel=0$. Using Eq.~\ref{eq:deltaf}, we are able to find the optimal $(\mathcal{E},\mathcal{B})$ operating point across the degenerate and non-degenerate regimes. 

We search the $N=1$, $K_a=1$ ortho manifold over
$0\le\mathcal E\le1000$~V/cm for the Sr isotopologues and
$0\le\mathcal E\le500$~V/cm for the Ra isotopologues, with
$0\le\mathcal B\le100$~G in all cases. We restrict our focus to
$|\Delta M|\le2$, corresponding to transitions accessible with one- or two-photon driving, with higher-$|\Delta M|$ pairs treated briefly in Appendix~\ref{ap:dm-ladder}. Transitions and $(\mathcal{E},\mathcal{B})$ points of interest are identified where $f_\mathcal{E}\approx f_\mathcal{B}\approx0$ and $\Delta\Sigma\ge0.1$. For each species, we identify $\mathcal{O}(10)$ transitions with $\mathcal{F}\gtrsim2$ for the strength of field noise we consider. The specific yield varies between different species, owing to variations in level structure density and dipole moments. More statistics on the yield of field-insensitive transitions are given in Appendix~\ref{ap:clock-search}.

The entries of Table~\ref{tab:magic} consist of the leading zero-$g$ factor pair and finite-field transition with the highest figure-of-merit, $\mathcal{F}$, for each isotopologue. The coherence time is calculated using experimentally relevant values for quasi-static noise of the lab-frame fields, with magnitudes $\delta \mathcal{E}=10$~mV/cm and $\delta \mathcal{B}=100$~\textmu G, split along parallel and perpendicular directions. These noise figures are relaxed by an order of magnitude from prior work~\cite{TakahashiEngineering2023}. Figure~\ref{fig:2D-magic} overlays the values of $\tau_\text{EM}$ obtained for all transitions found for these noise figures. 

\begin{figure}[tbp!]   
\centering
\includegraphics[width=\columnwidth]{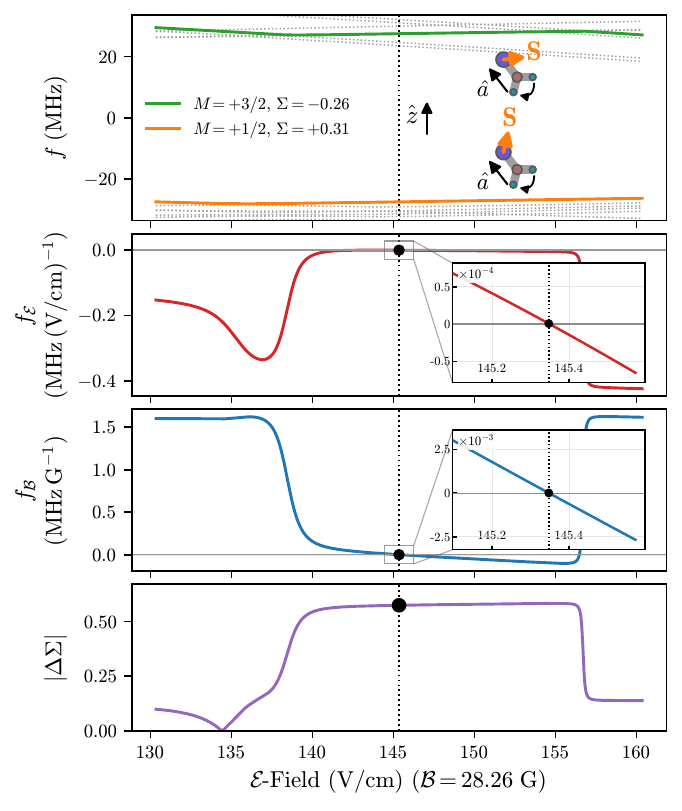} 
\caption{Engineered clock transition in $^{88}$Sr$^{14}$NH$_2$: the pair $M(+\tfrac{1}{2} \to +\tfrac{3}{2})$ at $\mathcal{B} = 28.26$~G.  From top: energies of the states in the transition (colored solid line), among the neighboring levels of both $M$ blocks (grey dashed lines). The molecule diagrams draw $\mathbf{S}$---$\hat{a}$ angles determined from the wavefunction composition. The clock point is given by the vertical dotted line. Next are plots of the differential Stark slope $f_\mathcal{E}$, the differential Zeeman slope $f_\mathcal{B}$, and the differential EDM sensitivity $|\Delta\Sigma|$. Both slopes vanish at $\mathcal{E}^{*} = 145.35$~V/cm, where $f = 55.0$~MHz, $|\Delta\Sigma| = 0.57$ and $\tau_{\rm EM} = 65$~s under the noise budget of Table~\ref{tab:magic}. Shading marks the clock criteria $|f_\mathcal{E}| < 10^{-3}$~MHz$\,$(V/cm)$^{-1}$, $|f_\mathcal{B}| < 10^{-3}$~MHz$\,$G$^{-1}$, $|\Delta\Sigma| > 0.1$.}
\label{fig:magic-crossing}
\end{figure}

The field-insensitive state pairs can be categorized into two groups: zero $g$-factor pairs of $\pm M$ states at $\mathcal{E}\neq0,\mathcal{B}\rightarrow0$~\cite{AndereggQuantum2023}, and finite frequency ($f\neq0$) clock transitions at nonzero fields: $\mathcal{E}\neq0, \mathcal{B}\neq0$~\cite{TakahashiEngineering2023,takahashi2025engineeredmolecularclocktransitions}. We find both types of field-insensitive transitions occur generally, even when effective Hamiltonian parameters are varied. The $\pm M$ pairs achieve $f_\mathcal{E}=0$ with time-reversal symmetry and $f_\mathcal{B}=0$ by tuning $\mathcal{E}$ through an avoided crossing of spin-rotation states. Meanwhile, the finite-field $\mathcal{B}\neq0$ transitions occur at the intersection of the contours of the $f_\mathcal{E}=0$ and $f_\mathcal{B}=0$ conditions. While exact first-order cancellation occurs only at a point, the exact nature of the crossing contours can open up a neighborhood of field values with sufficiently suppressed first-order sensitivity. A representative finite-field zero and its differential Stark, Zeeman, and eEDM
responses are shown in Fig.~\ref{fig:magic-crossing}.

The two transition types offer complementary operating modes. The zero $g$-factor $\pm M$ pairs enable standard spin-precession-style experiment protocols~\cite{AndereggQuantum2023}, while the $B\neq0$ transitions are more similar to experiments with clocks~\cite{takahashi2025engineeredmolecularclocktransitions} and qubits. In either case, the superposition state can be prepared with standard tools, such as microwave/RF pulses, two-photon Raman transitions~\cite{takahashi2025engineeredmolecularclocktransitions}, or STIRAP~\cite{ACMECollaborationImproved2018}. After the free evolution time $\tau$, the superposition accumulates a $T$-odd, orientation-odd phase, $|\psi(\tau)\rangle = \tfrac{1}{\sqrt{2}}\left(e^{-i\Delta\phi_{ij}}|i\rangle + |{j}\rangle\right)$, which is mapped onto populations with another unitary pulse and read out projectively. For these transitions, first-order field noise is subleading, and the residual differential sensitivities set the coherence times, shown in Table~\ref{tab:magic}.

These $\mathcal{B}\neq0$ transitions have already been demonstrated in
$^{174}$YbOH~\cite{takahashi2025engineeredmolecularclocktransitions}, where
frequency switching between a clock transition and a field-sensitive
transition at a single $(\mathcal{E},\mathcal{B})$ point corrects systematic
errors. Our 2D maps (Figure~\ref{fig:2D-magic}) extend this to asymmetric molecules and show the additional complexity provides many opportunities for measurements. Of additional interest are regions where field-insensitive ridges of two clock transitions intersect. Comparing transitions at such an intersection would cross-check the EDM signal channel, since a genuine EDM-induced energy shift $\delta f$ must scale with the EDM sensitivity $\delta f_1 / \Delta \Sigma_1 = \delta f_2 / \Delta \Sigma_2$ of each transition, while field-induced systematics generally do not. 

\subsection{Optical Trap Shifts}

\begin{figure*}[htpb]
    \centering
    \includegraphics[width=\linewidth]{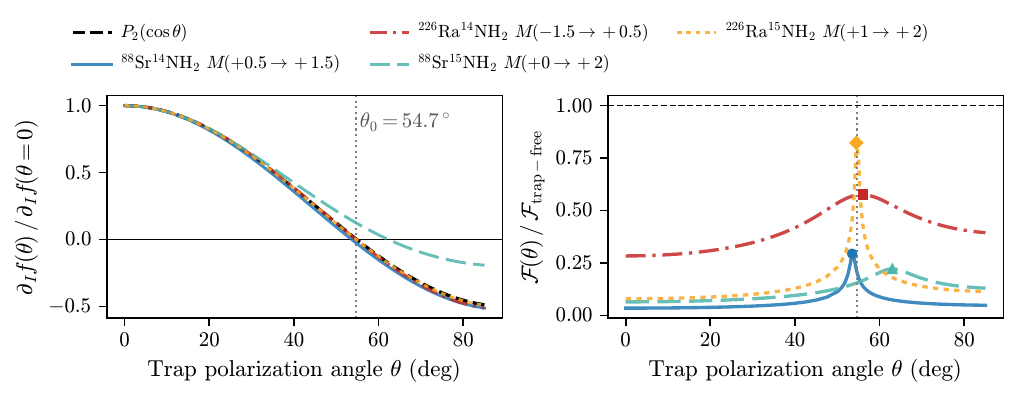}
    \caption{Trap-tilt tuning of the finite-field clock transitions of
Table~\ref{tab:magic}, one per isotopologue.
Left:~The slope of frequency vs trap intensity, $\partial f/\partial I \equiv \partial_I f$ follows $P_2(\cos\theta)$ (dashed) and vanishes near the magic angle $\theta_0$. The $^{88}$Sr$^{15}$NH$_2$ pair finds its magic angle instead near $62.5^\circ$, owing to a cancellation between the tensor $p=0$ term and second-order shifts from $|p|=1,2$ terms.
Right:~Recovery of the figure of merit $\mathcal{F}$, normalized by its trap-free value, at a $1$~MHz trap depth with $1\%$ intensity noise. Each transition is
evaluated at its in-trap optimum point $(\mathcal{E}^*,\mathcal{B}^*)$ with $\theta$ scanned, and the markers give the optimal angle $\theta^{*}$ that recovers measurement sensitivity. }
    \label{fig:magic-angle}
\end{figure*}

To further extend EDM measurement bounds, next-generation neutral molecule experiments aim to use optically trapped molecules with second-scale coherence~\cite{Kozyryev2017PolyEDM,HutzlerSearches2020}. However, the focused trapping light introduces perturbations that must be accounted for in any precision measurement~\cite{Burchesky2021Rotational,Bause2025,robichaud_parity-doublet_2026}. The two primary effects are ac Stark shifts between measurement states, and inelastic scattering of molecules from the off-resonant trapping light. The inelastic scattering rate can be suppressed by operating with large detunings and shallow traps. As with trapped diatomic and triatomic molecules, scattering lifetimes of $\mathcal{O}(10-100)$~seconds are expected in $\sim$1~MHz deep optical traps at $1064$~nm~\cite{Bause2025}. Conversely, the effect of differential ac Stark shifts will depend more strongly on the molecule's rotational structure and the choice of $(\mathcal{E},\mathcal{B})$ operating point. In this section, we investigate the effect of optical trapping on the rotational states of ATMs in detail, and show that engineered EDM transitions can be made robust against trap shifts. 

The ac Stark effect is set by the dynamic polarizability $\bm{\alpha}(\omega)$ of the molecule, and the coupling to the trap field can be represented using spherical tensor notation as~\cite{Caldwell2020Sideband,Bause2025}
\begin{equation}
    H_{\rm ac} = -\tfrac{1}{4}E_{\text{trap}}^2 \sum_{k=0,1,2}\sum_{p} (-1)^p\, T^k_p(\hat{\bm{\varepsilon}},\hat{\bm{\varepsilon}}^*)\, T^k_{-p}(\bm{\alpha}(\omega)).
\end{equation}
Here, $E_{\text{trap}}$ is the trap field amplitude, $T^k_p(\hat{\bm{\varepsilon}},\hat{\bm{\varepsilon}}^*)$ is a lab-frame component of the rank-$k$ spherical tensor formed by the polarization unit vector $\hat{\bm{\varepsilon}}$, which couples to $T^k_{-p}(\bm{\alpha})$, the corresponding lab-frame projection of the molecule's dynamic polarizability at the trap frequency.

The polarizability tensor is defined in the molecule frame and is diagonal for a $C_{2v}$ molecular structure (i.e., $\alpha_{aa},\alpha_{bb}, \alpha_{cc}$), similar to the moment of inertia tensor. Full calculation of the dynamical polarizability is left for future work. Here, we instead perform a simple estimate of the scale of the polarizability tensor to enable a representative study of trap shifts and develop general schemes for their mitigation. Additionally, we consider only the case of linear trapping light polarization, which causes the $k=1$ term in $H_{ac}$ to vanish by symmetry. We focus primarily on the $k=2$ tensor term, as the $k=0$ scalar term is isotropic and its effects are common-mode. 

The primary difference in $H_{\rm ac}$ between ATMs and the linear diatomic and triatomic molecules considered previously~\cite{Caldwell2020Sideband, Bause2025} is that ATMs generically break cylindrical symmetry $(\alpha_{bb}\neq\alpha_{cc})$, resulting in two sources of tensor polarizability: the usual ``axial'' term, $T^2_{q=0}(\alpha)\propto 2\alpha_{aa}-\alpha_{bb} - \alpha_{cc}$, as well as an additional asymmetry term $T^2_{q=\pm2}(\alpha)\propto\ \alpha_{bb} - \alpha_{cc}$. This additional term connects states with $\Delta K=\pm 2$ and can serve as a resource for state engineering. Further details of the polarizability tensor estimate are given in Appendix~\ref{sec:supp-polarizability}.

The tensor-trap sensitivity of time-reversed pairs of $\pm M$ states tuned to a ``zero $g$-factor'' crossing has previously been studied experimentally with CaOH~\cite{AndereggQuantum2023} and theoretically with BaOH~\cite{Bause2025}. Following a similar approach, we take $\mathcal{E}$, $\mathcal{B}$, and $\hat\epsilon$ collinear, such that $M$ remains a good quantum number and the $|p|=1,2$ spherical tensor components of $H_\text{ac}$ vanish. For a degenerate zero-$g$-factor pair, the collinear geometry retains a $p=0$ tensor that is $M$-even by time-reversal symmetry. For $\pm M$ states near $B\sim 0$, the differential ac Stark shift therefore vanishes by symmetry rather than by field tuning. 

Two residual sources of sensitivity remain in the collinear case. The first is sensitivity to transverse couplings from the $|p|=1,2$ components of the trap light due to polarization tilt and ellipticity, as well as stray transverse static fields. These interactions connect the $\pm M$ states through $2|M|$ steps of coupling via intermediate sublevels. For time-reverse pairs separated by $\Delta M=2$, the resulting sensitivity is quadratic in the transverse fields and can be suppressed by the application of a small longitudinal bias $\mathcal{B}$-field~\cite{AndereggQuantum2023,Bause2025}. However, for $M=\pm1/2$ pairs, there are no intermediate coupling steps and the transverse sensitivity is linear, requiring a larger bias field. We note that pairs with $\Delta M>2$ have even higher-order transverse sensitivities and require correspondingly smaller bias $\mathcal{B}$-fields. While we do not consider them in detail here, as they would require multiple pulses of one- or two-photon couplings for state preparation and readout, the $\Delta M>2$ pairs are enumerated in Appendix~\ref{ap:dm-ladder}.

We find that the application of a bias $\mathcal{B}_\text{bias}$-field to suppress transverse couplings for near-degenerate $\pm M$ states can perturb the $f_\mathcal{E}\approx f_\mathcal{B}\approx 0$ condition. The non-zero $f_\mathcal{BB}$ value means the bias field restores a fraction of the first-order sensitivity to parallel field noise, and the coherence time falls as $\tau\propto 1/\mathcal{B}_\text{bias}$. For $M=\pm1/2$ zero-$g$ pairs in $^{14}$N isotopologues, the effect is significant. Operation at the optimized 10~mG bias field value caps the coherence time near 2~s (Table~\ref{tab:magic}), far below the value without transverse field noise. These pairs remain useful, as every improvement in magnetic field noise translates to an improvement in coherence time. Nonetheless, among the candidate states, the $M=\pm 1$ pairs obtained from $^{15}$N isotopologues demonstrate superior noise properties. Additionally, zero $g$-factor states with $\Delta M\ge3$ will have further higher-order noise sensitivity, at the expense of requiring multi-photon addressing. 

The robustness of $M=\pm1$ compared to $M=\pm1/2$ pairs can be connected to the choice of nitrogen isotope used in the ligand. Since the more abundant $\mathcal{M}$--$^{14}$NH$_2$ isotopologue with $I_N=1$ has half-integer grand total angular momentum $F$, there is no protection from mixing of $M=\pm 1/2$ pairs. The $\Delta M\ge 3$ pairs are more robust to mixing and noise, but they require multi-photon addressing for preparation and readout. Meanwhile, $F$ is integer for the $\mathcal{M}$--$^{15}$NH$_2$ isotopologue with $I_N=1/2$. As a result, these species generally exhibit $M=\pm 1$ pairs with favorable properties and conventional preparation and readout methods. 

For $B\neq0$ EDM clock transitions between non-degenerate states, we are no longer protected from tensor trap shifts by time-reversal symmetry. In experiments with optically trapped molecules, tensor shifts are instead often managed with ``magic angle'' tuning~\cite{Burchesky2021Rotational,robichaud_parity-doublet_2026}. The trap polarization $\hat \epsilon$ is tilted by an angle $\theta$ with respect to the bias magnetic field $\mathcal{B}$. With the quantization axis along $\mathcal{B}$, the rank-2 polarization tensor is $T^2_{p=0}(\hat{\bm{\epsilon}},\hat{\bm{\epsilon}}^*)\propto 3\cos^2{\theta} -1$, which vanishes at the magic angle $\theta_0=\arccos{\sqrt{1/3}}\approx 54.7^\circ$. However, $\theta\neq0$ also introduces off-diagonal couplings through the $|p|=1,2$ components of $H_{\rm ac}$. These can be suppressed by operating with sufficient bias field magnitude, $\mu_B\mathcal{B}\gg \langle H_{\rm ac}\rangle$, such that the Zeeman splittings of neighboring $M$ levels detune the tensor trap couplings. The trap angle can additionally be tuned to a specific optimum $\theta^*\neq\theta_0$, such that the residual first-order $p=0$ shift cancels against the second-order $|p|=1,2$ shifts. The magic angle is generally intensity dependent, as the second-order shifts are only zeroed at a specific intensity, and residual quadratic trap shifts remain. While this limitation can be overcome with ``magic wavelength'' trapping~\cite{RuttleyLonglived2025}, this requires a more detailed understanding of the excited-state landscape than what is currently known. 

For the finite-field clock transitions of Table~\ref{tab:magic}, we find that an optimal $\theta^*$ often exists for a given transition and trap intensity, as illustrated in Fig.~\ref{fig:magic-angle}. The existence of such an optimal angle is largely independent of the uncertainties in the polarizability tensor estimate (see Appendix~\ref{sec:supp-polarizability}), and we can generally recover at least $\mathcal{O}(0.1)$ or more of the trap-free figure-of-merit. Figure-of merit analysis in the presence of the trap are provided in Appendix~\ref{ap:trap-noise}. The cancellation of tensor shifts near the magic angle is found to vary in sharpness for various transitions, but generally requires trap angle stability better than $0.1^{\circ}$. These observations are consistent with the comparison of energy scales: the trap is weaker than the Stark, Zeeman, and spin--rotation interactions responsible for tuning the electron-spin and molecular-dipole orientations, so the trap angle tunes the perturbation without significantly changing the transition character.

\section{Discussion and Outlook}

We have shown that $C_{2v}$ asymmetric top molecules, specifically alkaline-earth monoamides, are a promising platform for next-generation searches for symmetry-violating physics. These molecules occupy a useful middle ground between structural simplicity and quantum-state complexity. Their near-symmetry supports tractable structure and enables optical-cycling, while their explicit asymmetry provides long-lived opposite-parity $K$-doublets within the vibronic ground state. Taking into account the suppression of relaxation from nuclear spin-statistics, we estimate a radiative lifetime far longer than any other experimental timescales (see Appendix~\ref{sec:lifetime} for details). This is in contrast to linear polyatomic molecules utilizing vibrationally excited states with $\ell$-doublets, where the observed coherence time is already larger than the radiative lifetime ~\cite{robichaud_parity-doublet_2026}. Moreover, the long-lived science states and long coherence times shown here are necessary for any scheme seeking to utilize entanglement to obtain a quantum advantage~\cite{ZhangQuantumEnhanced2023,Salas-EstradaNondestructive2026}. Together with recently demonstrated 1-D laser cooling of CaNH$_2$~\cite{li2026photoncyclinglasercooling}, C$_{2v}$ asymmetric top molecules represent the next generation of trapped-molecule precision measurements.

We have also provided a theoretical and computational framework for understanding and modeling the detailed rotational, spin-rotational, and hyperfine structure of $C_{2v}$ asymmetric top molecules. Using relativistic \textit{ab-initio} calculations, we have predicted effective Hamiltonian constants required to model the field-tuning, EDM sensitivity, and state lifetime of these molecules. Additionally, we find that electron-correlation effects are especially important for the calculation of the $\epsilon_{aa}$ spin-rotation tensor in heavy species, observing deviations from Curl-type relations particularly in RaNH$_2$. Further investigations are needed to understand the nature of the relativistic effects contributing to $\epsilon_{aa}$, and separately to calculate additional properties of interest, such as the polarizability tensor, possibly enabling magic wavelength trapping~\cite{RuttleyLonglived2025}.

This work extends prior investigations into engineering states and transitions for EDM measurements~\cite{AndereggQuantum2023,TakahashiEngineering2023,takahashi2025engineeredmolecularclocktransitions}, both toward new molecular systems, and also toward addressing perturbations from optical trapping. Similar to diatomic and triatomic systems, we find that asymmetric top molecules generically exhibit many field-insensitive transitions. We provide a quantitative characterization of these transitions' transverse field noise, and show it becomes the limiting factor for $\Delta M=1$ zero $g$-factor states in half-integer $F$ molecules. We also perform a detailed study of ac Stark shifts from optical trapping light, at an angle $\theta$, in the presence of bias $\mathcal{E}$ and $\mathcal{B}$ fields. We show that for EDM-sensitive clock transitions, magic angle tuning of the trap polarization can be used to eliminate first-order shifts from trap intensity noise, recovering long interrogation times with realistic experimental conditions. The opportunity to utilize both zero $g$-factor states and finite $B\neq0$ transitions provides a variety of possible measurement protocols that can be adapted to specifics of molecular structure or the experimental setup. 

The mechanism shown here, using fields to dress states and engineer transitions, is essentially the static-field counterpart of dynamical decoupling protocols. In dynamically dressed systems, oscillating control fields perform time-averaging of noise and modify the effective coupling of a transition to environmental perturbations, i.e., tuning a $g$-factor to zero~\cite{Timoney2011}. In our case, the dressing interactions are the static Stark and Zeeman fields acting through the molecule's intrinsic avoided crossings arising from spin-rotation and hyperfine structure. In the future, it is possible that both approaches could be complementary and used together to further improve measurement protocols. 

The combination of long lifetime, engineered coherence protection, and a heavy nucleus in a polar molecule leads to significant projected measurement reach. The statistical sensitivity for the eEDM follows from the shot-noise limit for a coherence time-limited repetition rate, $\delta d_e = \hbar/(W_d|\Delta\Sigma|\sqrt{\tau N_p T})$, with the electronic enhancement factor $W_d$ coupling to the eEDM, coherence time $\tau$, $N_p$ molecules per cycle, and total integration time $T$. Calculated values of $W_d$ are available for SrNH$_2$, BaNH$_2$, YbNH$_2$, and RaNH$_2$ in Ref.~\cite{frenettVibrationalBranchingFractions2024}. Using $W_d=110$~GV/cm for RaNH$_2$, along with $ \langle\Delta \Sigma\rangle = 0.8$, $N_p=10^3$, $\tau=10$~s, and $T=1$~week, we obtain a projected sensitivity of $\delta d_e \approx 1\times10^{-31}\;e\,\text{cm}$, more than an order of magnitude beyond the current best statistical sensitivity of $2\times 10^{-30}\; e\,\text{cm}$~\cite{Roussyimproved2023}. This limit could be further improved by utilizing techniques for capturing and trapping more molecules~\cite{YoussefEnhanced2026,NasirOrder2026}, multiplexing measurements, and eventually through advanced quantum metrology techniques~\cite{ZhangQuantumEnhanced2023,Salas-EstradaNondestructive2026}. Additionally, any such measurement with $\tau=10$~sec will require suppression of blackbody radiation effects with a cryogenic environment. 

We would also like to point out another distinct feature of $K$-doublet that provides an advantageous knob for other types of precision measurement: tunable parity splitting. In the rotational Hamiltonian, the term $(B-C)/4$ couples $\Delta K = 2$, which gives rise to an energy gap between parity states. For a near prolate top like $\mathcal{M}$--NH$_2$ molecules, the asymmetry factor is small, and the parity energy gap decreases sharply and monotonically with increasing $K$. On the other hand, by increasing $N$, the parity gap is made larger. Selection of $K$-doublets with a tunable energy gap can be useful in various settings, including the following examples: \textbf{(i) Trapped-ion symmetry-violation searches}. The JILA eEDM search~\cite{Roussyimproved2023} utilizes a small $\Omega$-doublet of HfF$^+$, set by electronic structure, for polarizability at a moderate $\mathcal{E}$-field. The geometric $K$-doublet is a tunable design parameter that can be chosen to match an accessible drive frequency. By selecting the metal center, an $\mathcal{M}$--NH$_2^+$ ion can be tailored to either sector: a paramagnetic center retains the unpaired electron for an eEDM search, while a heavy octupole-deformed nucleus targets the nuclear sector. \textbf{(ii) Parity-violation searches}. The same field-tunable opposite-parity $K$-doublets are also attractive for nuclear-spin-dependent and nuclear-spin-independent parity-violation (PV) searches. The PV interactions typically manifest in $\Sigma$-like states but are suppressed in the $\Delta$ states whose $\Omega$-doublets are exploited. In $^2\Sigma$ diatomics such as BaF, however, the opposite-parity levels are a full rotational quantum apart, so tuning a PV-sensitive pair to degeneracy requires fields around 4600~G~\cite{altuntas2018}. On the other hand, the $\ell$-doublets of excited-bending-mode linear polyatomics were proposed to reach the crossing at fields roughly two orders of magnitude smaller~\cite{norrgard_nuclear-spin_2019}. A geometric $K$-doublet offers the same small-field advantage in both electronic and vibrational \emph{ground} states, and provides tunability with variable $K$-manifold.

The properties and generality of the system allow for metrological manipulation of entangled species for EDM measurements ~\cite{ZhangQuantumEnhanced2023}. A heavy spinful nucleus combined with the NH$_2$ ligand extends the platform to the nuclear sector, enabling magnetic quadrupole moments with $^{173}$Yb ($I=5/2$) and Schiff moments with $^{225}$Ra ($I=1/2$)~\cite{PhysRevA.100.032514,PhysRevLett.76.4316,YuProbing2021}. Although isotopes with a spinful heavy nucleus contain additional hyperfine structure, the same numerical framework can search their enlarged spectra for simultaneous zeros of the differential Stark and Zeeman slopes, as demonstrated for other nuclear-spin systems in Ref.~\cite{TakahashiEngineering2023}. Extending our calculations to these odd isotopes (and to YbNH$_2$, which is expected to be perturbed by low-lying electronic states) is left to future work. Together, these results suggest that ATMs, and in particular alkaline-earth monoamides, are promising candidates for next-generation BSM searches using precision molecular spectroscopy.

\section*{Acknowledgment}
We thank Phelan Yu and Timothy C. Steimle for helpful discussions. We thank Roy Ready, John M. Doyle, Grace K. Li, Alex Frenett, Lo{\"i}c Anderegg, and Benjamin L. Augenbraun for their feedback on the manuscript. Y.Y. is supported by the W.~M.~Keck Foundation. The authors thank the Center for Information Technology of the University of Groningen for their support and for providing access to the Hábrók high-performance computing cluster.
The work of I.A.A. was partially supported by the project \textit{Probing Particle Physics with Polyatomic Molecules}, project number OCENW.M.21.098 of the research program M2, which is financed by the Dutch Research Council (NWO).
L.F.P. acknowledges the support from project number VI.C.212.016 of the talent program VICI, financed by the Dutch Research Council (NWO), and support from the Scientific Grant Agency of the Slovak Republic (project 1/0254/24).

\appendix

\section{Long-lived $K$-doublets and finite-temperature lifetime}
\label{sec:lifetime}

\subsection{Radiative Lifetime}

\subsubsection{$K_a=1$ States}

The $K_a=1$ science manifold is protected against spontaneous rotational decay by nuclear-spin statistics.  Within the ground vibronic state, anti-symmetry under exchange of the two protons correlates the ortho manifold ($I_H=1$) with odd $K_a$ and the para manifold ($I_H=0$) with even $K_a$. Further symmetry classification details are given in Appendix~\ref{ap:MS_group}. An ortho---para transition therefore involves $\Delta K=1$, which changes the $(12)\equiv C_2$ symmetry properties of the rotational representation: $\Gamma_{\rm rot} \in \{B_1,B_2\}\rightarrow \{A_1,A_2\}$. To maintain the Pauli exchange symmetry of the total representation $\Gamma_{\rm tot}=\Gamma_{\rm rot}\otimes\Gamma_{\rm nuc}$, the nuclear spin factor of an ortho---para transition operator must also be odd under the $(12)$ exchange operation. For two identical spin-$1/2$ nuclei, such an $(12)$-odd operator is rank-1 and proportional to $\mathbf{I}_\Delta=\mathbf I_{H,1}-\mathbf I_{H,2}$, whereas $\mathbf{I}_H=\mathbf I_{H,1}+\mathbf I_{H,2}$ and $\mathbf I_{H,1}\!\cdot\!\mathbf I_{H,2}$ are $(12)$-symmetric and diagonal in $\mathbf{I}_H$. 

For the low-$K_a$ levels of near-prolate $C_{2v}$ molecules, the leading $(12)$-odd contribution to the hydrogen hyperfine Hamiltonian is the off-diagonal dipolar component $T_{ab}$:

\begin{equation}
    T_{ab}[(I_{\Delta})_a S_b+(I_\Delta)_bS_a] = \mathbf{I}_{\Delta}\cdot \frac{\mathbf{T}_1-\mathbf{T}_2}{2}\cdot \mathbf{S}. \label{ap:eq:Tab}
\end{equation}
where $T_{ab}=\frac{1}{2}\left(\mathbf{T}_1-\mathbf{T}_2\right)_{ab}$. 

The \textit{ab-initio} calculations in section~\ref{sec:ab-initio-calc} reveal $T_{ab}$ to be $\sim$1~MHz. This is much smaller than the reported 58 MHz for a 90-degree bend angle of NH$_2$~\cite{EngelsStudy1992}, primarily due to the fact that the unpaired valence electron in $\mathcal{M}$--NH$_2$ is centered around the polarized metal atom $\mathcal{M}^+$. 

Consider the transition $K_a = 1\rightarrow0$, which is nominally forbidden as shown above. $T_{ab}$ admixes a fraction $c_{p} = T_{ab}/\Delta E_{K_a}$ of a nearby para level into the ortho state, with selection rules $\Delta K_a=\rm odd$ and $\Delta K_c =\rm even$. Using $N_{K_aK_c}$ notation, we have for the dressed wavefunction, 
\begin{equation}
    |\widetilde{N_{1 K_c}}\ket \approx |N_{1K_c}\ket|I_H=1\ket+\sum_{N^\prime} c_{p}(N,N^\prime)  |N^\p_{0 N^\prime}\ket|I_H=0\ket,
\end{equation}
where the sum is over $N^\prime$ states mixtures with weight $c_p(N,N^\prime)$, which includes the $N,N^\prime$ dependence in the $T_{ab}$ transition matrix element and in the $\Delta E_{K_a}$ gap. We have also simplified $K_c^\p=N^\p$, valid for $K_a=0$. These $c_p$ admixtures allow the $K_a=1$ decay to borrow intensity from the E1 dipole-allowed $a$-type transition, which has selection rules $\Delta K_a=0$ and $|\Delta K_c|=1$. 

The resulting ortho-para decay rate for $K_a=1, N $ is:

\begin{widetext}
\begin{equation}
    A_{o\to p}(N_{1K_c})= \sum_{N^\p,N^\pp,K_c^\pp}\left|c_{p}(N,K_c;N^\p)\right|^2 \frac{16 \pi^3 f_{o\to p}^3 D_a^2}{3\varepsilon_0 h c^3}S_{\rm rot}(N^\p, K_c ; N^\pp,K_c^\pp)
\end{equation}
\end{widetext}
where $D_a$ is the $a$-axis electric dipole moment, $S_{\rm rot} (N^\p,K_c;N^\pp,K_c^\pp)$ is a rotational transition matrix element, typically $\mathcal{O}(1/3)$, and the transition frequency for the radiative decay  is $f_{o\to p}$. 

For the $1_{1K_c}$ estimate, we consider a representative case with $T_{ab}=1$~MHz, $A=390$~GHz, $(B+C)/2=6$~GHz, $B-C=0.1$~GHz, and $D_a=2$~D. The rank-2 $T_{ab}$ tensor has non-zero matrix elements $c(1,1)\approx1.4\times10^{-6}$ and $c(1,3)\approx1.4\times 10^{-6}$ for $K_c=1$, while $c(1,2)\approx2\times10^{-6}$ for $K_c=0$. The total admixtures for both $1_{10}$ and $1_{11}$ are very similar at $|c_{p}|^2 \approx 4 \times 10^{-12}$, and summing over final states gives similar values for $1_{11}$ and $1_{10}$ of $A_{o\to p}(1_{10/11}) \approx 7.7\times 10^{-15}$~Hz. This corresponds to an estimated ortho-para lifetime of approximately 4 million years.

Given this long lifetime, it becomes important to consider the possibility of coherence loss by decay \textit{within} $N=1,K_a=1$. Specifically, the upper parity doublet state, $1_{10}$, has an E1-allowed decay pathway to the lower parity doublet state, $1_{11}$. For a representative $100$~MHz parity splitting, the decay rate becomes $A_{E1}(1_{10})=2.3\times10^{-14}$~Hz, which is $3\times$ larger than the $A_{o\to p}(1_{10})$ rate. The combination of both rates sets the total lifetime for the $1_{10}$ state at 1 million years. 

While the decayed population remains in $K_a=1$, it has lost coherence, and therefore this lifetime bounds the coherence time achievable using $K_a=1$ doublets. Further, the application of electric fields will mix the doublets and can quench the lower doublet lifetime (if there remain lower-energy states left to decay to). Specifically, when polarized in an electric field, the parity doublets will split into three manifolds: an upper and lower Stark manifold describing oriented molecules with $\langle M_N K_a\rangle=\pm1$, and an intermediate manifold describing aligned molecules with $\langle M_N K_a\rangle=0$. The E1 decay satisfies $\Delta K_a=0$ and cannot directly couple the upper and lower Stark manifolds, which have $K_a=1$ and $K_a=-1$ for $\Delta M_N=0$. However, the E1 decay will be allowed from the upper oriented manifold to the intermediate aligned manifold, which consists of mixtures of $K_a=\pm 1$, as well as from the intermediate manifold to the lower oriented manifold of states. 

Our E1-borrowing model for ortho-para transitions is supported by the recent observation of an ortho-para transition in the $C_{2}$ system S$_2$Cl$_2$~\cite{PhysRevLett.119.173401}. There, the ortho-para states are mixed by the off-diagonal nuclear quadrupole tensor $\chi_{ab}$ instead of $T_{ab}$. The key difference is that S$_2$Cl$_2$ has $\sim$GHz scale separations of ortho and para levels, which enhances the intensity borrowing admixture $c_{p}$. Using their observed ortho/para intensity ratio of $10^{-3}$ with our estimates, we recover their reported lifetime of a few thousand years.

\subsubsection{$K_a=2$ States}

Because $K_a=2$ and $K_a=0$ have the same nuclear-spin symmetry,
their coupling requires no ortho-para mixing. Instead,
the $\Delta K=\pm2$ asymmetry terms in the rotational Hamiltonian mix the $K_a=2$ and $K_a=0$ states. Considering the $N=2$ states, for example, the asymmetry mixing only occurs between the positive parity $2_{02}$ state in $K_a=0$ and the $2_{20}$ state in $K_a=2$. While the negative parity $2_{21}$ state in $K_a=2$ cannot mix with $K_a=0$, it will end up having a similar lifetime due to the $K_a=0$ state being mixed. Defining $\bar B=(B+C)/2$, the matrix element and mixing coefficients for $2_{20}$ are
\begin{align}
    c_{\mathrm{asym}}
    &= \frac{\langle 2_{02}|H_{\rm rot}|2_{20}\ket}{4(A-\bar B)}
    =\frac{\sqrt{3}(B-C)}{4(A-\bar B)}.
\end{align}
To first order, the admixed wavefunctions can be written
\begin{align}
    |\widetilde{2_{20}}\rangle
    &\simeq
    |{2_{20}}\rangle+c_{\mathrm{asym}}|2_{02}\rangle,\\
    |\widetilde{2_{02}}\rangle
    &\simeq
    |{2_{02}}\rangle-c_{\mathrm{asym}}|2_{20}\rangle.
\end{align}
For the positive parity $2_{20}$ state, the E1 transition couples the $+c_{\rm asym}|2_{02}\ket$ admixture to the negative parity states $1_{01}$ and $3_{03}$. We note there will be interference effects from $3_{03}$ itself being mixed with $3_{21}$. Meanwhile, for the negative parity $2_{21}$ state, the E1 transition couples to the ground $2_{02}$ state via intensity borrowing through the $-c_{\rm asym}|2_{20}\rangle$ admixture. After computing the relevant matrix elements and summing over final states, we find both parity states have the same effective line strength to first order, $S_{\rm rot}=2/3$. 

For a representative estimate, we again take $A=390$~GHz, $(B+C)/2=6$~GHz, $B-C=0.1$~GHz, and $D_a=2$~D, obtaining $|c_{\rm asym}|^2\approx 1.3\times10^{-8}$. The resulting decay rate for $N=2$, $K_a=2$ manifold is $A_{\rm asym}(2_{20/21})= 1.4\times 10^{-9}$ Hz, corresponding to a $K_a=2$ lifetime of 23 years.

Finally, we also estimate the E1-allowed decay between the two $K_a=2$ parity doublets. The first-order mixing $c_{\rm asym}$ is responsible for parity doubling effects at second-order, connecting $K_a=2$ and $K_a=-2$ via intermediate $K_a=0$ states. Considering only the rotational Hamiltonian, the parity splitting is given by $|c_{\rm asym}|^2 \times4(A-\bar B)\approx 20~\text{kHz}$. As expected, this is much smaller than the $K_a=1$ doublets, and we can neglect this decay channel due to the $\omega^3$ suppression of the decay rate. 

\subsubsection{Field-Induced Quenching}
For a comprehensive lifetime estimate, we also consider Stark and/or Zeeman quenching from tuning and mixing of levels. At first order, neither field can flip $I_H$ from singlet to triplet. Further, for $N =1, K_a = 1$ states, the ortho-para splitting is $\sim$390~GHz, and using field tuning to close the gap and induce ortho-para mixing would require $\mathcal{E}\gtrsim400$~kV/cm or $B\gtrsim10$~T. Therefore, for low-lying rotational states, field quenching is negligible. We mention that much higher $N$ states belonging to different $K_a$ stacks can have accidental near degeneracies, but we do not consider them further here. 

At higher order, there exist $\mathcal{B}$-field-induced effects that couple to $I_\Delta$ and are odd under $(12)$ exchange. Specifically, the magnetic susceptibility of the molecule includes the effect of the magnetic shielding tensors~\cite{brownRotationalSpectroscopyDiatomic2003} $\mathbf{\sigma}$ of the two protons in the NH$_2$ ligand. The interaction Hamiltonian is given by
\begin{equation}
    \sigma_{ab}\left[\mathcal{B}_a (I_\Delta)_b + \mathcal{B}_b (I_\Delta)_a\right]= \boldsymbol{\mathcal{B}}\cdot \frac{(\boldsymbol{\sigma}_1-\boldsymbol{\sigma}_2)}2\cdot(\mathbf{I}_{H,1}-\mathbf{I}_{H,2}),
\end{equation}
which is written in the same form as Eq.~\ref{ap:eq:Tab} to illustrate the antisymmetric character of $\sigma_{ab}$. For an order of magnitude estimate, we calculate NMR shielding tensors, obtaining an off-diagonal component $\sigma_{ab}\sim5$~ppm. The interaction scale with the proton magnetic moment is given by $g_p \mu_N \, \mathcal{B} \, \sigma_{ab} \approx \mathcal{B}[\text{T}]\times  200~\text{Hz}$, orders of magnitude weaker than $T_{ab}$ at 1~T applied field.

Finally, the M2 magnetic gradient interaction can also couple to $I_\Delta$ via the transition magnetic-quadrupole moment. Intuitively, the magnetic gradient in the molecule frame can distinguish between the two protons and flip spins independently. Defining the proton distances along the $\hat{b}$-axis as $\boldsymbol{\delta}_1$ and $\boldsymbol{\delta}_2$, the M2 interaction Hamiltonian is
\begin{equation} 
g_p\mu_N\,(\nabla\mathcal{B})_{ab}  (I_\Delta)_{a}  \, \delta_b =\mathbf{I}_{\Delta}\cdot \boldsymbol{\nabla\mathcal{B}}\cdot\frac{(\boldsymbol{\delta}_1-\boldsymbol{\delta}_2)}{2}
\end{equation}
Using a representative value of $\delta_b \approx 0.8$~\AA, we would require a gradient of $|\nabla B|\sim 3\times 10^6$~T/cm to reach the scale of $T_{ab} \sim 1$~MHz.

\subsection{Blackbody Lifetime}

Blackbody radiation (BBR) in the electric-dipole approximation preserves nuclear-spin symmetry and keeps the ortho and para stacks separate. Direct BBR-driven transitions within the $K$-doublet are also negligible because the transition-rate factor $\omega^3\bar{n}(\omega,T)$ strongly suppresses transitions at kHz--MHz frequencies. For the molecules considered here, the dominant room-temperature BBR loss instead arises from excitation of low-lying vibrational modes. Using calculated vibrational frequencies and transition dipole moments, Ref.~\cite{VilasBlackbody2023} predicts a 300-K ground-state lifetime of $1.7$~s for CaNH$_2$, with an estimated uncertainty of approximately $20\%$. Rescaling the mode-resolved rates by the Planck occupation factors at 77~K gives a lifetime of order $10^3$~s. Population that subsequently returns to the science manifold has lost its phase coherence, so BBR-induced escape places an upper bound on the achievable internal-state coherence time.

BBR pumping effects on lifetime have been observed directly in trapped molecules. Ref.~\cite{VilasBlackbody2023} extracts a ground-vibrational-state BBR lifetime for optically trapped CaOH molecules of $\tau_{\mathrm{BBR}}=1.3^{+0.3}_{-0.2}$~s. Room-temperature BBR-driven rotational pumping of trapped OH and OD gives corresponding lifetimes of $2.8$ and $7.1$~s, respectively~\cite{HoekstraOptical2007}. Cooling the trap environment to 17~K subsequently enabled OH confinement for nearly one minute by suppressing both BBR pumping and collisions with residual background gas~\cite{HaasLongterm2019}. Cryogenic operation has likewise extended the rotational-state lifetimes of a polar molecular ion by approximately an order of magnitude relative to room temperature~\cite{ChaffeeHighFidelity2025}. High-optical-access cryogenic platforms have also been demonstrated for neutral-atom tweezer arrays, and suppression of BBR-induced Rydberg transitions continues as ongoing work~\cite{zhangHighOpticalAccess2025}. These experiments establish cryogenic shielding as a practical route to suppressing BBR-induced loss and decoherence in trapped atoms and molecules.

\section{Clock-transition search and electromagnetic dephasing}
\label{ap:clock-search}

\begin{table}[!htpb]
\centering
\caption{Cumulative figure-of-merit distribution of the transitions found with $|\Delta M|\leq2$ and $\mathcal{F}\geq0.25$, where $\mathcal{F}=|\Delta \Sigma|\sqrt{\tau_{\rm EM}}$. Points are obtained after transverse noise evaluation and
noise-averaged re-optimization of the $(\mathcal{E,B})$ point. }
\label{tab:catalog-yield}
\begin{ruledtabular}
\begin{tabular}{lrrrr}
Species & $\mathcal F\ge0.25$ & $\mathcal F\ge1$ &
$\mathcal F\ge2$ & $\mathcal F\ge5$\\
\hline
Sr$^{14}$NH$_2$ & 431 & 179 & 60 & 0\\
Sr$^{15}$NH$_2$ & 224 & 82 & 32 & 0\\
Ra$^{14}$NH$_2$ & 92 & 51 & 34 & 9\\
Ra$^{15}$NH$_2$ & 28 & 24 & 16 & 8\\
All & 775 & 336 & 142 & 17\\
\end{tabular}
\end{ruledtabular}
\end{table}

\begin{figure*}[!htpb]
\centering
\includegraphics[width=\textwidth]{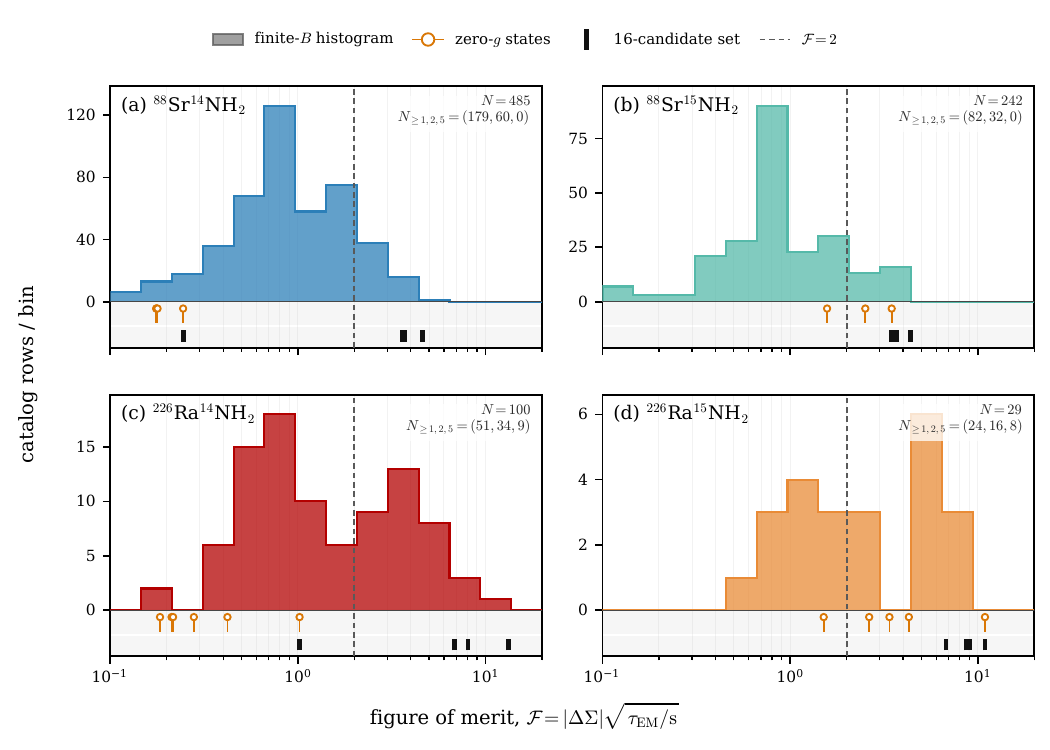}
\caption{Distribution of $\mathcal F$ for points found with $\mathcal F \geq0.1$.  (a-d) show the species considered. The histograms show the finite-field transitions binned logarithmically, and the 18 zero $g$-factor time-reverse pairs are shown below separately as open circles. Heavy marks below the histogram identify the transitions in Table~\ref{tab:magic} and Table~\ref{tab:additional-candidates}.}
\label{fig:fom-census}
\end{figure*}

\begin{figure*}[htpb!]
\centering
\includegraphics[width=\textwidth]{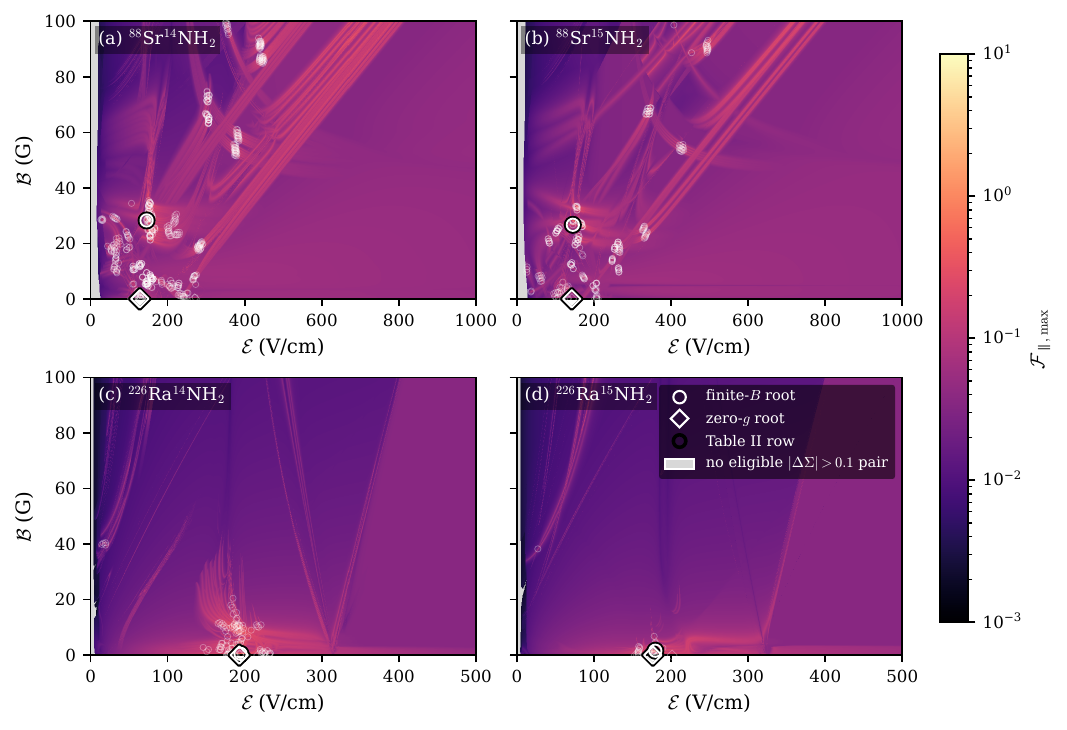}
\caption{2-D plots of the longitudinal figure of merit.  At each field point, the
color gives the maximum over catalog transitions with $|\Delta\Sigma|>0.1$
under longitudinal-only noise share of the Table~\ref{tab:magic} noise budget.  Circles mark finite-field transition points, diamonds mark zero $g$-factor points, and heavy outlines mark
the eight main-text headline transitions. One logarithmic
color scale is used for all four panels.}
\label{fig:eb-fom-map}
\end{figure*}

\begin{figure*}[!htpb]
\centering
\includegraphics[width=0.92\textwidth]{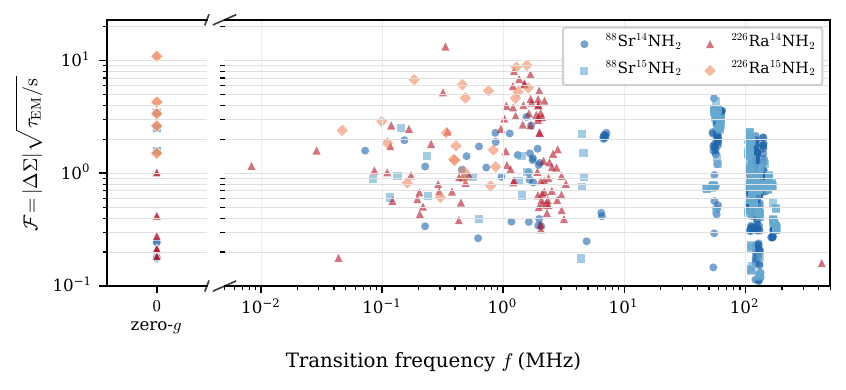}
\caption{Figure of merit $\mathcal{F}$ derived from longitudinal and transverse noise, plotted versus transition frequency.  The 18 zero $g$-factor pairs occupy a separate categorical
strip.  Color and marker shape identify the
isotopologue.}
\label{fig:fom-frequency}
\end{figure*}

\subsection{Frequency Broadening}
\label{ap:tauem}

We model the longitudinal electric and magnetic field noise as independent,
quasistatic (i.e., slower than internal molecular interactions) Gaussian random variables,
$\delta\mathcal E\sim\mathcal N(0,\sigma_{\mathcal E}^2)$ and
$\delta\mathcal B\sim\mathcal N(0,\sigma_{\mathcal B}^2)$.  Expanding the
transition frequency to second order gives
\begin{equation}
\begin{split}
\delta f={}&f_{\mathcal E}\delta\mathcal E+f_{\mathcal B}\delta\mathcal B\\
+{}&f_{\mathcal{EB}}\delta\mathcal E\delta\mathcal B
+\frac12f_{\mathcal{EE}}\delta\mathcal E^2
+\frac12f_{\mathcal{BB}}\delta\mathcal B^2,
\end{split}
\label{eq:df-expand}
\end{equation}
where all derivatives are evaluated at the $(\mathcal{E,B})$ operating point.  The variance is computed as $\operatorname{Var}\left[\delta f\right]=\langle \delta f^2\rangle-\langle\delta f\rangle^2$. First we consider the variance from independent parallel fluctuations $\delta \mathcal{E}_\parallel, \delta \mathcal{B}_\parallel$
\begin{equation}
\begin{split}
\operatorname{Var}_{\parallel} \! \left[\delta f\right]={}&
f_{\mathcal E}^2\sigma_{\mathcal E}^2
+f_{\mathcal B}^2\sigma_{\mathcal B}^2\\
+{}&f_{\mathcal{EB}}^2\sigma_{\mathcal E}^2\sigma_{\mathcal B}^2
+\frac12f_{\mathcal{EE}}^2\sigma_{\mathcal E}^4
+\frac12f_{\mathcal{BB}}^2\sigma_{\mathcal B}^4.
\end{split}
\label{eq:frequency-variance}
\end{equation}
Note that if there is non-zero co-variance, we would obtain additional terms proportional to the second derivatives. The term $f_\mathcal{E B}$ is particularly of importance as it scales any non-reversing correlated noise, which could provide a false EDM. Fortunately, the ability to test various magic and sensing transitions helps disentangle such systematics~\cite{takahashi2025engineeredmolecularclocktransitions}.

The total variance used in the main text Table~\ref{tab:magic} also contains the transverse
responses,
\begin{equation}
\operatorname{Var}_{\rm tot}\!\left[\delta f\right]=
\operatorname{Var}_{\parallel}\!\left[\delta f\right]
+\operatorname{Var}_{\delta\mathcal E_\perp}\![f]
+\operatorname{Var}_{\delta\mathcal B_\perp}\![f].
\label{eq:frequency-variance-total}
\end{equation}
We evaluate the transverse terms by direct sampling, which remains
well-behaved for degenerate time-reverse pairs, where the perturbative expression becomes singular as $\mathcal B\rightarrow0$. After transverse noise evaluation, we perform another re-optimization of the operating point. This is particularly important for the time-reverse zero $g$-factor pairs. 

\begin{table*}[!htpb]
\centering
\caption{Fractional contributions to the transition-frequency variance for the transitions enumerated in the main text and appendix: the eight operating points in
Table~\ref{tab:magic} and the eight additional candidates in
Table~\ref{tab:additional-candidates}.  $f_\mathcal{E}$ and $f_\mathcal{B}$ are the linear
longitudinal terms, $f_\mathcal{EE}$ and $f_\mathcal{BB}$ the diagonal curvatures, $f_\mathcal{EB}$ the mixed
Stark--Zeeman curvature, and $f_\mathcal{E_\perp}, f_\mathcal{B_\perp}$ the transverse terms.  Entries
below $0.1\%$ are shown as upper bounds.}
\label{tab:budget-shares}
{
\begin{tabular}{llrrr|rrrrrrr}
\hline\hline
Species & $M_i\to M_j$ & $\mathcal E^*$ & $\mathcal B^*$ & $\tau_{\rm EM}$
& $f_\mathcal{E}$ & $f_\mathcal{EE}$ & $f_\mathcal{EB}$ & $f_\mathcal{B}$ & $f_\mathcal{BB}$ & $f_\mathcal{E_\perp}$ & $f_\mathcal{B_\perp}$\\
& & (V/cm) & (G) & (s) & \multicolumn{7}{c}{Share of $\operatorname{Var}(\delta f)$ (\%)}\\
\hline
Sr$^{14}$NH$_2$ & $-\tfrac12\to+\tfrac12$ & 128.95 & 0.0104 & 2.0 & 33.4 & $<0.1$ & $<0.1$ & 42.1 & $<0.1$ & $<0.1$ & 24.4\\
Sr$^{14}$NH$_2$ & $+\tfrac12\to+\tfrac32$ & 145.35 & 28.26 & 65.1 & 9.7 & 42.4 & 15.3 & 0.2 & $<0.1$ & 32.4 & $<0.1$\\
Sr$^{14}$NH$_2$ & $+\tfrac12\to+\tfrac12$ & 146.01 & 28.54 & 41.8 & 0.1 & 77.8 & 4.6 & $<0.1$ & $<0.1$ & 17.5 & $<0.1$\\
Sr$^{14}$NH$_2$ & $-\tfrac12\to+\tfrac32$ & 167.23 & 24.95 & 35.5 & $<0.1$ & 20.9 & 2.4 & $<0.1$ & $<0.1$ & 76.7 & $<0.1$\\
Ra$^{14}$NH$_2$ & $-\tfrac12\to+\tfrac12$ & 193.64 & 0.0084 & 1.6 & 37.7 & $<0.1$ & $<0.1$ & 30.3 & $<0.1$ & $<0.1$ & 32.0\\
Ra$^{14}$NH$_2$ & $-\tfrac32\to+\tfrac12$ & 194.52 & 0.47 & 261.7 & $<0.1$ & 9.9 & 74.8 & $<0.1$ & $<0.1$ & 0.1 & 15.3\\
Ra$^{14}$NH$_2$ & $-\tfrac32\to-\tfrac12$ & 196.08 & 5.86 & 709.7 & 1.4 & 1.8 & 73.2 & $<0.1$ & $<0.1$ & 23.3 & 0.3\\
Ra$^{14}$NH$_2$ & $-\tfrac32\to-\tfrac32$ & 188.65 & 3.83 & 519.8 & 0.4 & 58.5 & 36.8 & $<0.1$ & $<0.1$ & 4.0 & 0.2\\
Sr$^{15}$NH$_2$ & $-1\to+1$ & 142.31 & $3{\times}10^{-5}$ & 711.0 & 15.1 & $<0.1$ & 70.0 & 14.8 & $<0.1$ & $<0.1$ & 0.1\\
Sr$^{15}$NH$_2$ & $0\to+2$ & 145.02 & 26.69 & 54.3 & $<0.1$ & 3.8 & 8.0 & $<0.1$ & $<0.1$ & 88.2 & $<0.1$\\
Sr$^{15}$NH$_2$ & $+1\to+2$ & 155.31 & 31.93 & 37.0 & $<0.1$ & 25.3 & 3.5 & $<0.1$ & $<0.1$ & 71.3 & $<0.1$\\
Sr$^{15}$NH$_2$ & $-2\to0$ & 145.82 & 28.31 & 38.7 & $<0.1$ & 44.8 & 3.8 & $<0.1$ & $<0.1$ & 51.3 & $<0.1$\\
Ra$^{15}$NH$_2$ & $-1\to+1$ & 176.37 & $5{\times}10^{-5}$ & 232.1 & 22.6 & $<0.1$ & 54.3 & 22.6 & $<0.1$ & $<0.1$ & 0.5\\
Ra$^{15}$NH$_2$ & $+1\to+2$ & 179.53 & 1.57 & 343.8 & 0.1 & 47.9 & 51.4 & $<0.1$ & $<0.1$ & 0.2 & 0.3\\
Ra$^{15}$NH$_2$ & $+1\to+1$ & 178.31 & 1.68 & 338.9 & 0.1 & 52.6 & 46.5 & 0.6 & $<0.1$ & 0.1 & 0.1\\
Ra$^{15}$NH$_2$ & $-1\to0$ & 176.37 & 0.0975 & 317.1 & 1.2 & 68.4 & 24.7 & 0.5 & $<0.1$ & $<0.1$ & 5.2\\
\hline\hline
\end{tabular}}
\end{table*}
\subsection{Transition Statistics and Noise Budgets}
\begin{table*}[!htpb]
\centering
\caption{Additional optimal finite-field transitions under the same longitudinal and transverse noise model as Table~\ref{tab:magic}.  The last column gives the coefficient that dominates $\operatorname{Var}[\delta f]$ at the quoted noise budget.
$f_{\mathcal{EE}}$ and $f_{\perp\mathcal E}$ are in
MHz/(V/cm)$^2$, and $f_{\mathcal{EB}}$ is in
MHz/[(V/cm)G].}
\label{tab:additional-candidates}
\begin{ruledtabular}
\renewcommand{\arraystretch}{1.2}
\begin{tabular}{llccccccc}
Species & Transition & $\mathcal E$ & $\mathcal B$ & $f$
& $\tau_{\rm EM}$ & $|\Delta\Sigma|$ & $\mathcal F$ & Limiting $\delta f$\\
& & (V/cm) & (G) & (MHz) & (s) & & &\\
\hline
Sr$^{14}$NH$_2$ & $M(+\tfrac12\to+\tfrac12)$ & 146.01 & 28.54 & 55.2 & 41.8 & 0.576 & 3.72
& $f_{\mathcal{EE}}=-6.0{\times}10^{-4}$\\
 & $M(-\tfrac12\to+\tfrac32)$ & 167.23 & 24.95 & 64.2 & 35.5 & 0.604 & 3.60
& $f_{\mathcal E\perp}=7.0{\times}10^{-4}$\\
Ra$^{14}$NH$_2$ & $M(-\tfrac32\to-\tfrac12)$ & 196.08 & 5.86 & 1.23 & 709.7 & 0.303 & 8.07
& $f_{\mathcal{EB}}=2.4{\times}10^{-3}$\\
 & $M(-\tfrac32\to-\tfrac32)$ & 188.65 & 3.83 & 1.35 & 519.8 & 0.301 & 6.86
& $f_{\mathcal{EE}}=4.2{\times}10^{-5}$\\
Sr$^{15}$NH$_2$ & $M(+1\to+2)$ & 155.31 & 31.93 & 59.3 & 37.0 & 0.607 & 3.69
& $f_{\mathcal E\perp}=6.5{\times}10^{-4}$\\
 & $M(-2\to0)$ & 145.82 & 28.31 & 55.3 & 38.7 & 0.590 & 3.67
& $f_{\mathcal E\perp}=5.2{\times}10^{-4}$\\
Ra$^{15}$NH$_2$ & $M(+1\to+1)$ & 178.31 & 1.68 & 1.28 & 338.9 & 0.475 & 8.74
& $f_{\mathcal{EE}}=6.1{\times}10^{-5}$\\
 & $M(-1\to0)$ & 176.37 & 0.0975 & 0.184 & 317.1 & 0.380 & 6.77
& $f_{\mathcal{EE}}=-7.4{\times}10^{-5}$\\
\end{tabular}
\end{ruledtabular}
\end{table*}

We search the $N=1$, $K_a=1$ ortho manifold in finite steps over 
$0\le\mathcal E\le1000$~V/cm for Sr$^{14,15}$NH$_2$ and $0\le\mathcal E\le500$~V/cm for Ra$^{14,15}$NH$_2$, and $0\le\mathcal B\le100$~G for all isotopologues.  The primary catalog is
restricted to $|\Delta M|\le2$, corresponding to one- or two-photon addressing.
Candidate clock transitions occur where $f_{\mathcal E}$ and
$f_{\mathcal B}$ are simultaneously below a threshold, while $|\Delta\Sigma|>0.1$. For every candidate point with $\mathcal{F}\geq 0.25$, we further refine the $(\mathcal{E,B})$ grid with progressively smaller step sizes until convergence. Each final point is a true root of $(\partial_\mathcal{E} f)^2+(\partial_\mathcal{B} f)^2=0$. 

Table~\ref{tab:additional-candidates} gives two additional finite-field
transitions for each isotopologue.  Together with the zero-$g$ factor and
finite-field entries in Table~\ref{tab:magic}, these are the best candidates found. Additionally, we provide fractional broadening from individual noise contributions in Table~\ref{tab:budget-shares}. Figure~\ref{fig:fom-census} shows the distribution of $\mathcal{F}$ across the points investigated. Figure~\ref{fig:eb-fom-map} maps the best figure of merit for all transitions over the $(\mathcal{E},\mathcal{B})$ plane, showing the regions where $f_\mathcal{E}=f_\mathcal{B}=0$. The slopes visible in the 2-D map are given by $M_N/M_S$, and indicate paths where electric and magnetic tuning compensate each other. Meanwhile, Figure~\ref{fig:fom-frequency} visualizes the figure of merit plotted against transition frequency. Operating points are abundant rather than finely tuned (Table~\ref{tab:catalog-yield}).

\subsection{Extension beyond $|\Delta M|\le2$}
\label{ap:dm-ladder}

\begin{table*}[ht!]
\centering
\caption{Time-reversed $\pm M$ pairs with no $\Delta M$ restriction, evaluated with the longitudinal and transverse field noise model from Table~\ref{tab:magic}.  For
$\Delta M\ge3$, $\mathcal B<0.01$~mG indicates only a small bias field and electric field detuning is needed to protect against transverse field noise.}
\label{tab:dm-operating}
\begin{ruledtabular}
\begin{tabular}{llrrrrr}
$M$ & Species & $\mathcal E^*$ (V/cm) & $\mathcal B_{\rm op}$ (mG)
& $\tau_{\rm EM}$ (s) & $|\Delta\Sigma|$ & $\mathcal F$\\
\hline
$\pm\frac12$ & Ra$^{14}$NH$_2$ & 193.64 & 8.36 & 1.6 & 0.816 & 1.02\\
$\pm\frac12$ & Sr$^{14}$NH$_2$ & 128.95 & 10.41 & 2.0 & 0.173 & 0.25\\
$\pm1$ & Ra$^{15}$NH$_2$ & 176.37 & 0.0457 & 232.1 & 0.717 & 10.93\\
$\pm1$ & Sr$^{15}$NH$_2$ & 142.31 & 0.0328 & 711.0 & 0.130 & 3.48\\
$\pm\frac32$ & Ra$^{14}$NH$_2$ & 196.15 & $<0.01$ & 303.7 & 0.826 & 14.39\\
$\pm\frac32$ & Sr$^{14}$NH$_2$ & 130.52 & $<0.01$ & 146.0 & 0.408 & 4.93\\
$\pm2$ & Ra$^{15}$NH$_2$ & 177.90 & $<0.01$ & 303.7 & 0.794 & 13.84\\
$\pm2$ & Sr$^{15}$NH$_2$ & 102.37 & $<0.01$ & 184.5 & 0.323 & 4.39\\
$\pm\frac52$ & Ra$^{14}$NH$_2$ & 200.06 & $<0.01$ & 304.8 & 0.831 & 14.51\\
$\pm\frac52$ & Sr$^{14}$NH$_2$ & 131.89 & $<0.01$ & 137.8 & 0.516 & 6.06\\
\end{tabular}
\end{ruledtabular}
\end{table*}

If the $|\Delta M|\le2$ restriction is lifted, more zero $g$-factor transitions are available with improved $|\Delta \Sigma|$ and further reduced transverse coupling sensitivity. In Table~\ref{tab:dm-operating}, we show additional time-reverse state pairs through $\Delta M=5$ using the same longitudinal and transverse noise model optimization as Table~\ref{tab:magic}.  A transverse
magnetic interaction changes $M$ by one unit per application, so coupling $-M$
to $+M$ first appears at order $2|M|=\Delta M$ and the low-field splitting
scales as $\mathcal B_\perp^{\Delta M}$.  

All evaluated $\Delta M\ge3$ pairs optimize at $<10$~\textmu G bias while
retaining $99\%$ of the figure-of-merit.  In radium,
$\tau_{\rm EM}\simeq304$~s from $\Delta M=3$ to 5, so the higher $\Delta M$ gain in $\mathcal F$ comes primarily from larger $|\Delta\Sigma|$.  A transverse-electric sweep to 30~V/cm produced no resolved splitting for these tested pairs
($<10^{-12}$~MHz) at our level of numerical precision. Higher $\Delta M$ transitions thus provide stronger transverse noise protection at
the cost of multiphoton or multipulse preparation and readout.

\section{Trap-Induced Light Shifts and Polarizability Estimates}
\label{ap:trap}

\subsection{Dynamic Polarizability Estimate}
\label{sec:supp-polarizability}

Evaluating trap-induced (AC-Stark) shifts of the magic transitions requires the
molecule-frame dynamic polarizability tensor of the $\tilde{X}\,^2A_1$ state at
the trap wavelength. For a $C_{2v}$ molecule, the tensor is diagonal in the
inertial frame with three independent components $(\alpha_{aa}, \alpha_{bb},
\alpha_{cc})$. We quote the reduced form $\tilde{\alpha} = \alpha/(2hc\epsilon_0)$
in Hz/(W/cm$^2$). The scalar component
$\tilde{\alpha}^{(0)}=(\tilde{\alpha}_{aa}+\tilde{\alpha}_{bb}+\tilde{\alpha}_{cc})/3$ produces the common-mode scalar shift $-\tilde{\alpha}^{(0)}I$,
where $I$ is the trap intensity.

We estimate the $\tilde \alpha$ components with a sum over the low-lying
metal-centered excited states,
\begin{equation}
\tilde{\alpha}_q(\omega) \;=\; \sum_n
\frac{2\,\Delta E_n\,|\langle \tilde{X}|d_q|n\rangle|^2}{\Delta E_n^{2}-\omega^{2}}
\;+\; \tilde{\alpha}_{\mathrm{core}},
\label{eq:sos}
\end{equation}
plus the closed-shell $\mathcal M^{2+}$ core polarizability
$\tilde{\alpha}_{\mathrm{core}}$~\cite{ChattopadhyayPolarizabilities2014}. The
molecule is well described as $\mathcal M^+$(NH$_2$)$^-$ with a single metal-centered
valence electron, so the sum is dominated by the three states derived from the
metal $np$ orbital. By $C_{2v}$ symmetry, the orbitals and therefore transition dipole moments are aligned to the principal axes, with the low-lying excited states corresponding to:
$\tilde{A}\,^2B_2$ (in-plane perpendicular $p$, $b$-polarized), 
$\tilde{B}\,^2B_1$ (out-of-plane $p$, $c$-polarized), and 
$\tilde{C}\,^2A_1$ ($p$ along the $\mathcal M$--N bond, $a$-polarized)
~\cite{BopegederaLaser1987,
MorbiHighres1998}. Each excited state contributes to its corresponding diagonal polarizability component $\alpha_{aa} (\tilde C), \alpha_{bb} (\tilde A), \alpha_{cc} (\tilde B)$. 

For SrNH$_2$ we use the measured band centers $\tilde{A} = 14274$, $\tilde{B} = 14724$, $\tilde{C} = 15862$~cm$^{-1}$\cite{BopegederaLaser1987}.  For unobserved RaNH$_2$, we use
the SOC-EOM-CCSD term energies 12515, 14547, and 15777~cm$^{-1}$ from
Ref.~\cite{zhangAnalyticGradientsRelativistic2023}.  The two calculated
$6d$-derived states are omitted because the leading $s\rightarrow d$
transition is E1 dipole forbidden, and molecular transition dipoles are
unavailable. Neglecting additional states in the sum underestimates the polarizability.  

No measured or \emph{ab-initio} electronic transition dipoles exist for any $\mathcal{M}$--NH$_2$ species, so we estimate them in the pure-precession limit~\cite{MorbiHighres1998} with atomic-like orbitals. We assign each
molecular $p$ projection the corresponding $\mathcal M^+$ $ns$--$np$ D-line strength
from the NIST Atomic Spectra Database \cite{NISTASD}. Comparisons with measured
and computed (relativistic-CCSD) polarizabilities and transition strengths in CaOH
\cite{HallasOptical2023,vilas_magneto-optical_2022} motivate an uncertainty of $\sim 20\%$ in the scalar polarizability and up to $50\%$ in the anisotropies. 

The spin-orbit interaction mixes the excited levels and redistributes the coupling strengths, particularly relevant when estimating the $T^2_{\pm2}(\alpha)$ anisotropy terms $\alpha_{bb}-\alpha_{cc}$.  For the $\tilde A/\tilde B$ pair, the ligand-field splitting competes with the off-diagonal spin--orbit interaction $a_{\rm SO}L_aS_a$
\cite{MorbiHighres1998,LiuRotational2018}. Diagonalizing this two-state
manifold gives $\Delta E_{\rm obs}=\sqrt{\Delta_{\rm el}^2+a_{\rm SO}^2}$ and $\tan(2\vartheta)=a_{\rm SO}/\Delta_{\rm el}$.  For equal radial transition dipoles, the mixing reduces the anisotropic term by
$\cos(2\vartheta)$. 

For SrNH$_2$, we use the measured SrOH value
    $a_{\rm SO}=263.5~\mathrm{cm}^{-1}$~\cite{PresunkaSrOH1994} and the observed
$\tilde A/\tilde B$ splitting of $\sim$$450~\mathrm{cm}^{-1}$
\cite{BopegederaLaser1987}.  These values give
$\Delta_{\rm el}\simeq365~\mathrm{cm}^{-1}$ and
$\cos(2\vartheta)\simeq0.81$.

For RaNH$_2$, all three metal-centered $p$ orbitals are mixed due to large spin-orbit coupling.  We use a simple
one-electron model
\begin{equation}
H_{\rm el}=
\operatorname{diag}(\epsilon_a,\epsilon_b,\epsilon_c)\otimes\mathbf 1_2
+\zeta_{\rm eff}\sum_{q=a,b,c}L_q\otimes\frac{\sigma_q}{2}.
\end{equation}
Here, $\epsilon_q$ are the pre-spin-orbit term energies of the orbitals split by the ligand field. For post-spin-orbit we use calculated SOC-EOM-CCSD term energies $(12515,14547,15777)~\mathrm{cm}^{-1}$ from Ref.~\cite{zhangAnalyticGradientsRelativistic2023}. To infer the orbital mixing,
we fix $\zeta_{\rm eff}=1540~\mathrm{cm}^{-1}$ and vary the pre-spin-orbit ligand-field energies
$\epsilon_q$ until the eigenvalues reproduce the theory values. The value $\zeta_{\rm eff}=1540~\mathrm{cm}^{-1}$ is estimated by scaling $\zeta_{\mathrm{Ra}^+}$ according to a reduction factor inferred from comparing CaOH, SrOH, and BaOH relative to their free-ion values
\cite{MorbiHighres1998,PresunkaSrOH1994,TandyBaOH2009,NISTASD}. In the end, we obtain the following state compositions:
\begin{equation}
\begin{array}{c|ccc}
E~(\mathrm{cm}^{-1}) & {\tilde A}~(\%) & {\tilde B}~(\%) & \tilde C~(\%)  \\
\hline
12515~(\tilde A) & 56 & 31 & 13 \\
14547~(\tilde B)  &  40 & 58 & 2 \\
15777~(\tilde C)& 4 & 11 & 85
\end{array}.
\end{equation}
The spin-orbit interaction significantly mixes the lower $\tilde A$ and $\tilde B$ states, while $\tilde{C}$ retains more of its character. The corresponding unperturbed energies are $(\epsilon_a,\epsilon_b,\epsilon_c)
=(15343,13452,14045)~\mathrm{cm}^{-1}$.

At 1064~nm, the dynamical polarizability estimate, in Hz/(W/cm$^2$), is
\begin{equation}
\begin{aligned}
(\tilde\alpha_{aa},\tilde\alpha_{bb},\tilde\alpha_{cc})_{\rm SrNH_2}
  &= (9.72,12.21,11.50),\\
(\tilde\alpha_{aa},\tilde\alpha_{bb},\tilde\alpha_{cc})_{\rm RaNH_2}
  &= (12.4,17.2,15.3),
\end{aligned}
\label{eq:alpha-estimates}
\end{equation}
with scalar mean values $\tilde\alpha^{(0)}$ of 11.14 and 14.97, respectively.  Isotopic substitution does not change the electronic polarizability at this level of estimate. 

Because the molecular polarizabilities are presently estimated rather than measured,
we tested the stability of the trap conclusions over a deliberately broad 18-point tensor grid. We held the scalar polarizability fixed, varied the axial anisotropy $T^2_0$ by $\pm20\%$,
and scaled the asymmetric anisotropy $T^2_{\pm 2}$ from $-1$ to $3$ relative to its estimated value. Throughout the grid of points, approximately $80\%$ of the high transitions studied retained a magic polarization angle at which the differential tensor light-shift vanishes. 

\subsection{Mitigating Trap Noise}
\label{ap:trap-noise}

We consider a linearly polarized 1064-nm optical dipole trap (ODT) with a 1~MHz depth and 1\% intensity noise resulting from finite molecule temperature in the trap.  For each finite-field transition candidate, we optimize with $H_{ac}$ over the operating parameters $(\mathcal E,\mathcal B,\theta)$.  To characterize angular tolerance, we then hold $(\mathcal E,\mathcal B)$ fixed and scan the polarization angle $\theta$ and evaluate the sensitivity to noise. 

\begin{figure*}[t]
\centering
\includegraphics[width=0.88\textwidth]{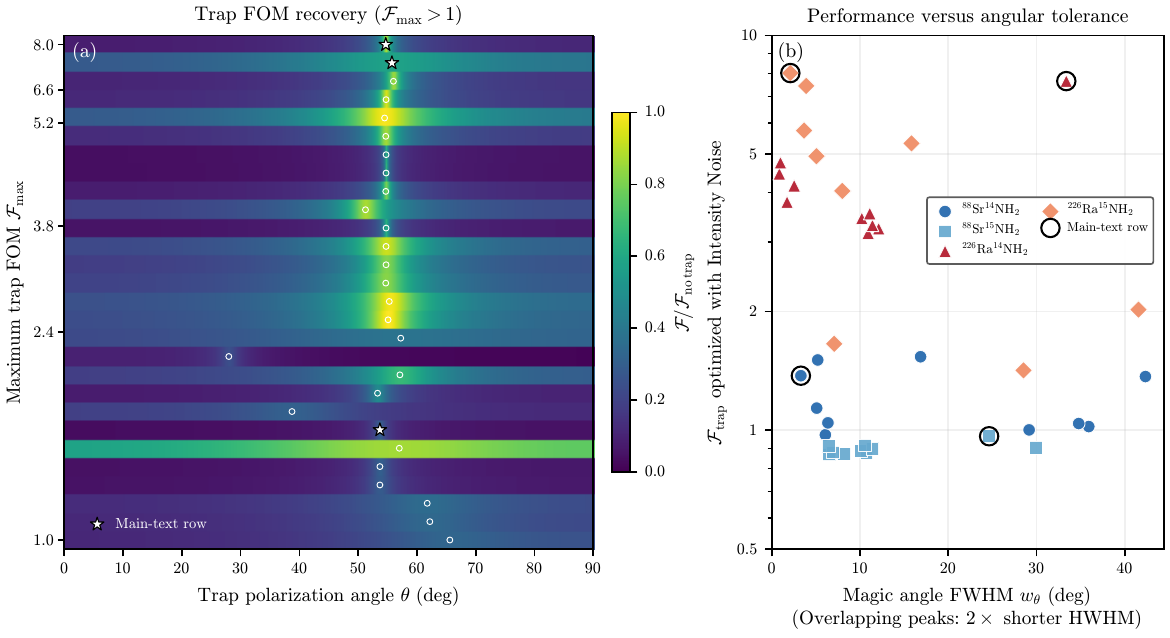}
\caption{Impact of optical trap polarization angle on the figure-of-merit $\mathcal F=|\Delta \Sigma|\sqrt{\tau/{\rm s}}$. We consider a 1-MHz deep, 1064-nm trap with 1\% intensity noise. The static
fields are held at each transition's trap-on optimum while the linear
polarization angle is scanned.  Left: recovery
$\mathcal F(\theta)/\mathcal F_{\rm no\,trap}$ for 28 transitions with
$\mathcal F_{\max}>1$, with rows ordered by $\mathcal F_{\max}$. Open circles mark the location of  $\mathcal{F}_{\max}$; stars identify main-text rows. Right: optimized
$\mathcal F_{\rm trap}$, including $1\%$
trap-intensity noise, plotted against angular width of the magic condition, for 39 transitions with
$\mathcal F_{\rm trap}>0.5$. }
\label{fig:magic-angle-ensemble}
\end{figure*}

We investigated the impact of the trap on the top 100 finite-field transitions,  selected with pre-trap figure of merit $\mathcal F \geq 2$. After reoptimizing $(\mathcal E,\mathcal B,\theta)$ with the trap present, 98 remained above $\mathcal F=0.25$, while two fell below this threshold. Figure~\ref{fig:magic-angle-ensemble} plots the impact of magic angle tuning for transitions that retain $\mathcal F_{\max}>1$ with the trap on. 

For the zero $g$-factor time-reversed pairs under collinear, linearly
polarized trapping light, the differential tensor shift vanishes by time-reversal
symmetry.  

\section{Adiabatic State Tracking}
\label{ap:tracking}
The dense production maps used for the transition catalog order states by energy at each grid point and evaluate point-local slopes, curvatures, and $\Sigma$ directly on the eigenstates, so they do not require global adiabatic labels. The tracking algorithm described here supplies the consistent state labels used for state-resolved maps and figures.

To extend the understanding of the system into the regime with both electric and magnetic fields, we need to perform high-fidelity adiabatic tracking of the level ordering in a two-dimensional $(\mathcal{E},\mathcal{B})$ grid. In principle, if each step imposed is reasonably small, the direction in which the algorithm follows should not matter. There are multiple ways to do the adiabatic tracking \cite{saberiAdiabaticTrackingQuantum2014,frameEigenvectorContinuationSubspace2018}. The most straightforward one would be to first track the $\mathcal{E}$-field progression, and then stack layers by incrementing the $\mathcal{B}$ field gradually. There is a caveat, which arises due to avoided crossings. The $\mathcal{E}$ and $\mathcal{B}$ fields break parity and time-reversal symmetry, respectively. Any tracking path aligned with a single field axis therefore samples a biased subset of crossings, and a single mislabeling propagates through every subsequent point along the sweep. We mitigate this problem using diagonal tracking through the ($\mathcal{E}$, $\mathcal{B}$) grid.

Constructing the Stark and Zeeman maps requires consistent eigenstate
labels across the $(\mathcal{E}_i, \mathcal{B}_j)$ grid, whereas direct
diagonalization returns eigenstates ordered by eigenvalue and therefore
scrambles wavefunction composition between neighboring states. We restore
consistent labels by adiabatically tracking states across the grid:
starting from the main diagonal and expanding outward, each newly
visited point $(i,j)$ is matched to its already-labeled neighbors $n$ by
minimizing the cost, with repeated $i,j$ summed over:
\begin{equation}
  C_{kk'} = \sum_{n} w_n\left[\alpha\left(E^{(n)}_{k}-E^{(ij)}_{k'}\right)^{2}
            - \beta\left|{\psi^{(n)}_{k}}{\psi^{(ij)}_{k'}}\right|
            \right],
  \label{eq:tracking_cost}
\end{equation}
where the sum runs over already-labeled neighbors $n$ of the point $(i,j)$,
$w_n = 1$ for nearest neighbors and $1/\sqrt{2}$ for diagonal neighbors,
and we take $\alpha = \beta = 1$ with energies in MHz. The first term penalizes eigenvalue (energy $E_i$) differences and the second rewards
high-fidelity behavior where state overlap $|\bra \psi_i|\psi_f\ket|$ is maximized. The optimal
relabeling permutation $\pi^{\ast} = \arg\min_{\pi}\sum_k C_{k,\pi(k)}$
is found by solving the linear assignment problem, after which
expectation values of all operators of interest are evaluated on the
relabeled grid. In high field, the spin components are entirely decoupled. In this regime, one need not worry about the avoided crossing-induced tracking mislabel. We can then assign and express the state with decoupled basis via Clebsch–Gordan coefficients:

\begin{widetext}
\begin{align}
|{N,K;(N,S)J,(J,I_N)F_N,(F_N,I_H)F,M}\ket &= \nonumber\\
\sum_{M_N,M_S,m_J,m_{I_N},m_{F_N},m_{I_H}}
&C^{J,m_J}_{N,M_N;S,M_S}\,
C^{F_N,m_{F_N}}_{J,m_J;I_N,m_{I_N}}\,
C^{F,M}_{F_N,m_{F_N};I_H,m_{I_H}} \nonumber\\
&\times |{N,K,M_N}\ket\otimes|{S,M_S}\ket\otimes|{I_N,m_{I_N}}\ket\otimes|{I_H,m_{I_H}}\ket
\end{align}
\end{widetext}

\begin{figure*}[htpb]
\centering
\includegraphics[width=\textwidth]{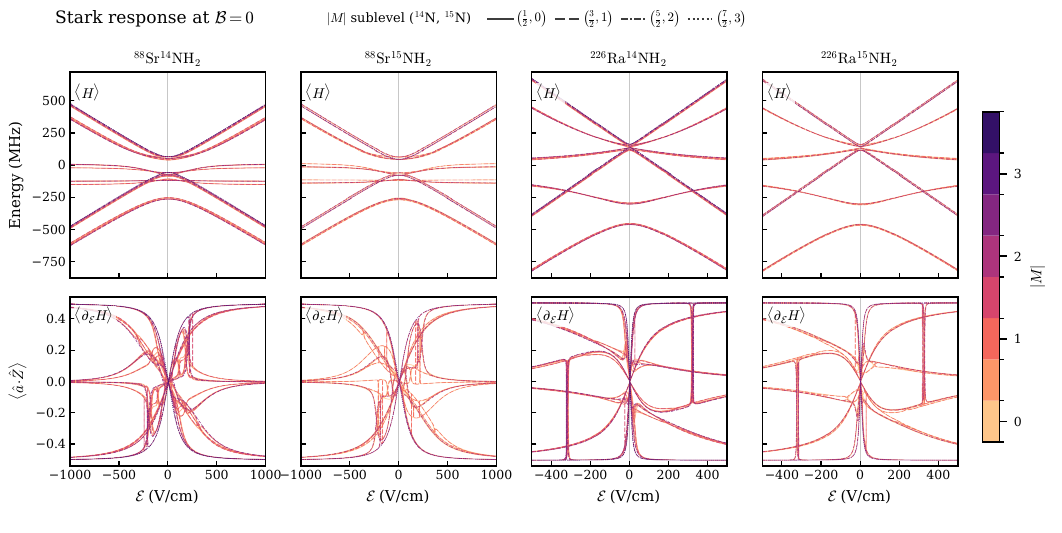}
\caption{Stark shifts of the $N=1$, $|K|=1$ manifold for all four
isotopologues at $\mathcal B=0$.  The upper row gives the state energies, and the lower row gives the body-axis orientation
$\langle\hat{a}\cdot\hat Z\rangle$.  Sr panels span
$|\mathcal E|\leq1000$~V/cm and Ra panels span
$|\mathcal E|\leq500$~V/cm.  Color identifies $|M|$, while
line style groups the four
$(|M|_{{}^{14}\mathrm{N}},|M|_{{}^{15}\mathrm{N}})$ rung pairs indicated above
the panels.}
\label{fig:stark-1d-four-species}
\end{figure*}

\begin{figure*}[htpb]
\centering
\includegraphics[width=\textwidth]{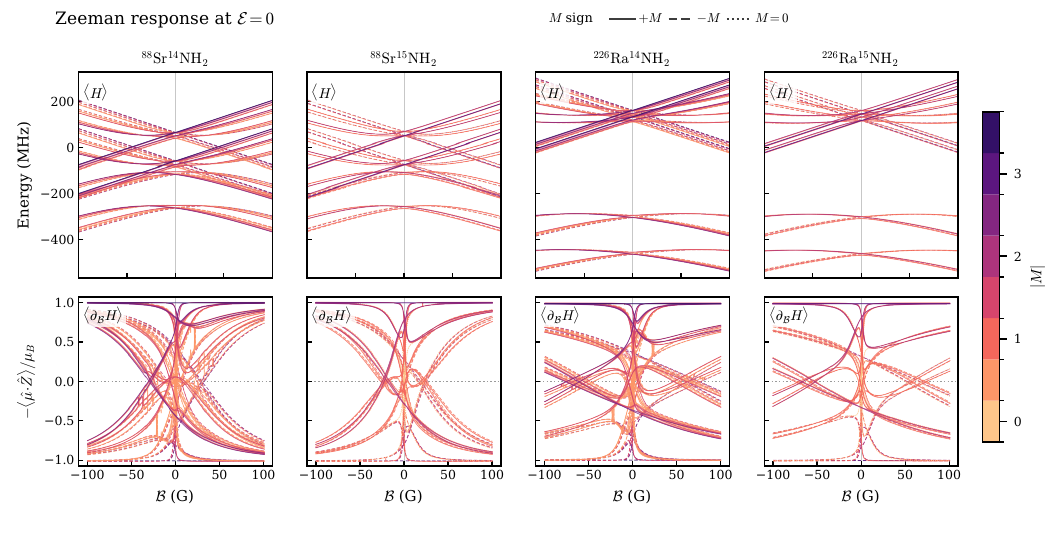}
\caption{Zeeman shifts of the same four manifolds at $\mathcal E=0$ over
$|\mathcal B|\leq100$~G. The upper row gives energies, and the lower row gives the dimensionless first-order magnetic sensitivity
$\partial_{\mathcal B}E/\mu_B
=-\langle\hat\mu\cdot\hat Z\rangle/\mu_B$ for each state.  The horizontal dotted reference marks zero first-order Zeeman slope.  Color identifies $|M|$; solid, dashed, and dotted state curves denote positive, negative, and zero $M$, respectively. }
\label{fig:zeeman-1d-four-species}
\end{figure*}

We show several example plots of molecules via this tracking method, see Figures~\ref{fig:stark-1d-four-species} and \ref{fig:zeeman-1d-four-species}. The several-hundred-MHz structure in both figures
reflects rotational asymmetry, spin-rotation, and hyperfine splittings.

\section{Molecular Structure and Group Theory}
\label{ap:MS_group}

\subsection{Rotational Structure}
 For the rotational states, we focus on the lowest rotational quantum numbers, $N=0,1,2,\ldots$, and ignore higher-order centrifugal corrections. The moment of inertia matrix,
\begin{align}
    I_{\alpha\beta}=\sum_i m_i \left(r_i^2\,\delta_{\alpha\beta} - r_{i,\alpha} r_{i,\beta} \right),
\end{align}
characterizes the molecule's rotational kinetic energy. The eigenvectors of $I_{\alpha\beta}$ define the principal axes of the molecular system, labeled by $a, b, c$. The rotational energies about each axis are characterized by the constants $A=\hbar^2/(2I_{aa})$, $B=\hbar^2/(2I_{bb})$, and $C=\hbar^2/(2I_{cc})$, with ordering convention $A > B > C$ ($I_{aa}<I_{bb}<I_{cc}$) \cite{bunkerMolecularSymmetrySpectroscopy1998,KochQuantum2019a}. 

In analogy to a classical rigid rotor, the rotational Hamiltonian is:
\begin{equation}
    H_\text{rot} = A \mathbf{N}_a^2 + B \mathbf{N}_b^2 + C \mathbf{N}_c^2  \label{eq:rot}
\end{equation}
where $\mathbf{N}_{a,b,c}$ are the rotational and electronic angular momentum (excluding spin) vectors along the principal axes of the molecule. As they are defined in the molecular frame, these operators have anomalous commutation relations, i.e., $[N_a,N_b] = -i \hbar N_c$, further detailed in \cite{zareAngularMomentumUnderstanding1988}.

We can provide a geometric picture of ATM dynamics by recasting Eq.~\ref{eq:rot} in terms of the perpendicular angular momentum $\mathbf{N}_\perp = \mathbf{N}_b+\mathbf{N}_c$ precessing about the $a$-axis with an azimuthal angle $\phi = \arctan{\mathbf{N}_c/\mathbf{N}_b}$. The rotational Hamiltonian becomes 
\begin{align}
    H_\text{rot} = A \mathbf{N}_a^2 + \frac{B+C}{2}\left(1+  \frac{B-C}{B+C}\cos{2\phi}\right)\mathbf{N}_\perp^2.
    \label{eq:rot2}
\end{align}
The average rotational constant $(B+C)/2$ sets the inertial energy scale for the precession of $\mathbf{N}_\perp$, while the rotational asymmetry $(B-C)/2$ generates a potential proportional to $\mathbf{N}_\perp^2\cos{2\phi} = \mathbf{N}_b^2-\mathbf{N}_c^2$, breaking cylindrical symmetry in the $bc$ plane and resulting in rotational parity doubling. For the molecules we consider here, since $B-C\ll B+C$, the potential in $\phi$ is weak compared to the kinetic energy, and the direction of $\mathbf{N}_\perp$ is delocalized \cite{KochQuantum2019a}.

The symmetric top wavefunctions $|N,K,M\ket$ describe the probability amplitude for finding the molecule frame ($xyz$) in a specific orientation with respect to the lab frame ($XYZ$). The rotations relating these two frames can be parameterized by the Euler angles ($\omega = \alpha,\beta,\gamma$), which define a conjugate angular position basis for the symmetric top wavefunctions:
\begin{align}
    \langle \omega | N, K, M\ket = \sqrt{\frac{2N+1}{8\pi^2}} \mathcal{D}^{(N)}_{M,K} (\omega)^*,
\end{align}
where $\mathcal{D}^{(N)}_{M,K}(\omega)$ is a unitary Wigner D-matrix. Ref.~\cite{littlejohnGaugeFieldsSeparation1997} further considers the position basis for molecular phase space.

In this paper, we consider ATMs near the prolate limit with $A\gg B\simeq C$, making the $a$-axis a natural choice for the molecular quantization axis $\hat{Z}$. Specifically, we identify the molecular quantization frame in terms of the principal axes according to the $I^r$ representation, $z=a$, $x=b$, $y=c$. With this choice, the rotational Hamiltonian can be written in a form that makes its structure and rotational symmetries more transparent:
\begin{align}
    H_\text{rot} &= A \mathbf{N}_z^2 + \frac{B+C}{2} \left(\mathbf{N}^2 - \mathbf{N}_z^2\right) +\frac{B-C}{4}\left(\mathbf{N}_+^2 + \mathbf{N}_-^2\right) \label{eq:1}\\
    &= \sum_{k=0,2} T^k(B_\text{rot}) \cdot T^k(\mathbf{N},\mathbf{N}). \label{eq:2}
\end{align}
Here, once again all $N_\alpha$ components are defined in the molecular frame, and in the second line we have written the Hamiltonian in compact spherical tensor form, with $T^k(B_\text{rot})$ representing the spherical tensors formed from the rotational constants $A,B,C$.  

The first two terms of Eq.~\eqref{eq:1}, which correspond to $T^0_0(N,N)$ and $T^2_0(N,N)$ in the spherical tensor expansion, conserve $K$ with the selection rule $\Delta K=0$, while the last term, corresponding to $T^2_{\pm 2} (N,N)$, mixes $K$ by coupling states separated by $\Delta K = \pm 2$, lifting the degeneracy of $\pm K$ states. This term is generated by the rotational asymmetry $B-C$ and is equivalent to the $\mathbf{N}_\perp^2\cos{2\phi}$ potential in Eq.~\eqref{eq:rot2}. In the molecules considered here, the rotational parity doubling is on the order of $(B-C)/2\simeq 10^2~\text{MHz}$, while the $\Delta K=2$ splitting is $4A(K+1)\simeq10^6~\text{MHz}$~\cite{frenettVibrationalBranchingFractions2024,ThompsenRotational2000}, which means the magnitude $|K|$ is conserved to $10^{-4}$ at the wavefunction level. Further, $N$ is still conserved, which restricts $K$ mixing to states with $|K|\leq N$. For example, in near-prolate tops considered here, the $N=1, K=\pm1$ parity doublet wavefunctions are analogous to the $N=1, \ell=\pm1$ states in linear bending modes that are actively used for precision measurements~\cite{Kozyryev2017PolyEDM,HallasOptical2023}.

\subsection{Parity States}

Due to rotational parity doubling and $\pm K$ mixing, ATM eigenstates for $K\neq0$ are given by symmetrized superpositions of $|N,\pm K,M\ket$ states, known as Wang functions~\cite{wangAsymmetricalTopQuantum1929}:
\begin{align}
    |N_{K_a,K_c},M\ket &:= \frac{1}{\sqrt{2}} \left( |N,K_a,M\ket + (-1)^{K_c} |N,-K_a,M\ket\right) \\
    &= \frac{1}{\sqrt{2}} \left( |N,K_a,M\ket \pm (-1)^{N-K_a} |N,-K_a,M\ket\right)
    \label{eq:wang}
\end{align}
The labels $K_a$ and $K_c$ describe the projection of $N$ on the symmetry axis in the prolate top limit ($A>B\simeq C$) or oblate top limit ($A\simeq B>C$). Because we are in the prolate limit, $K_a$ is well defined, and the sum $K_a + K_c$ is either $N$ or $N+1$, depending on the $K_c$ value, which is given by the parity of the wavefunction in Eq.~\ref{eq:wang}. 

The Wang functions can be classified by how they transform under the symmetries of $H_\text{rot}$. This classification enables the determination of rotational selection rules, parity symmetries, and nuclear spin-statistics. Specifically, the Wang functions transform as irreducible representations (irreps) of the group formed by $\pi$ rotations about the principal axes, $\{R^\pi_a,R^\pi_b,R^\pi_c\}$, and the identity, $E$. This constitutes the dihedral point group $D_2$, isomorphic to the Klein four-group $\mathbb{Z}_2\times \mathbb{Z}_2$, with elements $\left\{(+1,+1),(+1,-1),(-1,+1),(-1,-1)\right\}$. In the ATM case, these group elements correspond to $(-1)^{K_a}, (-1)^{K_c}$, denoting the even (e) or odd (o) parities of $K_a$ and $K_c$. In Table~\ref{tab:c2v-char}, the Wang functions are labeled by $(-1)^{K_a}, (-1)^{K_c}$ and classified as irreps of $D_2\simeq\mathbb{Z}_2\times \mathbb{Z}_2$. Geometrically, the nontrivial irreps can be identified with a double-headed arrow describing alignment of angular momentum along each axis. Such an object is invariant under a rotation about a collinear axis and reflects under $\pi$ rotations about perpendicular axes. 

The space-fixed parity operator, denoted $E^*$, reverses all lab-frame coordinates and is important for determining selection rules. The action of $E^*$ on the Euler angles is used to obtain the equivalent rotation describing the effect of $E^*$ in the molecular frame. For the $I^r$ representation, it can be shown that $R_y^\pi$ is the equivalent rotation to $E^*$, resulting in $K_a\xrightarrow{E^*} -K_a$ and  $K_c\xrightarrow{E^*} K_c$. The rotational parities of the Wang functions are presented in Table~\ref{tab:c2v-char}, showing that the space-fixed parity eigenvalue is given by $(-1)^{K_c}$.

\subsection{$C_{2v}$ Point Group}
\label{sec: group theory}
The rotational part of the asymmetric-top wavefunction is primarily parametrized by molecular rotation projections $K_a$ and $K_c$, where the subscript indicates the axis of reference. Here, the $C_{2v}$ group has four irreducible representations: $A_1, A_2, B_1, B_2$ and four operations: identity ($E$), $\pi$ rotation around the principal axis ($C_2[z]$), and planar reflections ($\sigma[xz]$, $\sigma[yz]$). We adopt the convention that $z = a$ is the principal axis. Geometrically, $C_2[z]$ tracks with $(-1)^{K_a}$, $\sigma[xz]$ tracks with $(-1)^{K_c}$, and $\sigma[yz]$ tracks with $(-1)^{K_a} \cdot (-1)^{K_c}$, leaving the following symmetry table:
\begin{table}[h!]
\centering
\caption{Mapping between $(\text{sgn}~K_a,\text{sgn}~K_c)$ and $C_{2v}$ irreps.}
\begin{tabular}{ccc}
\hline\hline
Symmetry & $(-1)^{K_a}$ & $(-1)^{K_c}$ \\ \hline
$A_1$ & $+$ & $+$ \\ 
$A_2$ & $+$ & $-$ \\ 
$B_1$ & $-$ & $-$ \\ 
$B_2$ & $-$ & $+$ \\ \hline\hline
\end{tabular}
\end{table}

Two identical spin-1/2 nuclei add to produce triplets and a singlet. These states are classified as either symmetric $(A_1)$ or antisymmetric $(B_2)$ under exchange $(C_2[z], \sigma[yz])$. The symmetric triplets are:
\begin{align}
\vert A_1, 1, 1 \rangle &= \left\vert \tfrac{1}{2}, \tfrac{1}{2} \right\rangle, \\
\vert A_1, 1, 0 \rangle &= \frac{1}{\sqrt{2}} \left( \left\vert \tfrac{1}{2}, -\tfrac{1}{2} \right\rangle + \left\vert -\tfrac{1}{2}, \tfrac{1}{2} \right\rangle \right), \\
\vert A_1, 1, -1 \rangle &= \left\vert -\tfrac{1}{2}, -\tfrac{1}{2} \right\rangle,
\end{align}
and the antisymmetric singlet is:
\begin{align}
\vert B_2, 0, 0 \rangle &= \frac{1}{\sqrt{2}} \left( \left\vert \tfrac{1}{2}, -\tfrac{1}{2} \right\rangle - \left\vert -\tfrac{1}{2}, \tfrac{1}{2} \right\rangle \right).
\end{align}

Two identical spin-1 nuclei add to produce quintets, triplets, and a singlet. These states are classified as either symmetric $(A_1)$ or antisymmetric $(B_2)$ under exchange $(C_2[z], \sigma[yz])$. The symmetric quintets are:
\begin{align}
\vert A_1, 2, 2 \rangle &= \vert 1, 1 \rangle, \\
\vert A_1, 2, 1 \rangle &= \frac{1}{\sqrt{2}} \left( \vert 1, 0 \rangle + \vert 0, 1 \rangle \right), \\
\vert A_1, 2, 0 \rangle &= \frac{1}{\sqrt{6}} \left( \vert 1, -1 \rangle + 2\vert 0, 0 \rangle + \vert -1, 1 \rangle \right), \\
\vert A_1, 2, -1 \rangle &= \frac{1}{\sqrt{2}} \left( \vert -1, 0 \rangle + \vert 0, -1 \rangle \right), \\
\vert A_1, 2, -2 \rangle &= \vert -1, -1 \rangle,
\end{align}
the antisymmetric triplets are:
\begin{align}
\vert B_2, 1, 1 \rangle &= \frac{1}{\sqrt{2}} \left( \vert 1, 0 \rangle - \vert 0, 1 \rangle \right), \\
\vert B_2, 1, 0 \rangle &= \frac{1}{\sqrt{2}} \left( \vert 1, -1 \rangle - \vert -1, 1 \rangle \right), \\
\vert B_2, 1, -1 \rangle &= \frac{1}{\sqrt{2}} \left( \vert 0, -1 \rangle - \vert -1, 0 \rangle \right),
\end{align}
and the symmetric singlet is:
\begin{align}
\vert A_1, 0, 0 \rangle &= \frac{1}{\sqrt{3}} \left( \vert 1, -1 \rangle - \vert 0, 0 \rangle + \vert -1, 1 \rangle \right).
\end{align}

\begin{table}
    \caption{$C_{2v} (M)$ character table describing the action of group operations (columns) on the irreducible representations (rows). The operations are: the identity $E$, exchanging identical nuclei $(12)$, space-fixed inversion (parity) $E^*$, and the combination of exchange and parity $(12)^*$. In addition to the MS group, we show the equivalent $C_{2v}$ point group operations for vibronic coordinates and the equivalent $D_2$ rotational operations on the molecular frame Euler angles. The column labeled Functions provides example classifications of linear, rotational, and quadratic degrees of freedom. We also show the classification of rotational eigenstates, labeled by the even (e)-odd (o) parity $K_a K_c$.}
    \label{tab:c2v-char}
    \resizebox{0.8\columnwidth}{!}{
    \begin{tabular}{c|cccc|c|c}
    \hline\hline
    $C_{2v}(M)$ & $E$ & $(12)$ & $(12)^*$ & $E^*$ & Functions & \\
    \hline
    $C_{2v}$ & $E$ & $C_{2}$ & $\sigma_{yz}$ & $\sigma_{xz}$ & & \\
    \hline
    $D_2$ & $R^0$ & $R^{\pi}_z$ & $R^{\pi}_x$ & $R^{\pi}_y$ & & $K_a K_c$ \\
    \hline
    $A_1$ & $+1$ & $+1$ & $+1$ & $+1$ & $z,x^2,y^2,z^2$ & $ee$ \\
    $A_2$ & $+1$ & $+1$ & $-1$ & $-1$ & $R_z, xy$ & $eo$ \\
    $B_1$ & $+1$ & $-1$ & $+1$ & $-1$ & $y, R_x, yz$ & $oo$ \\
    $B_2$ & $+1$ & $-1$ & $-1$ & $+1$ & $x, R_y, xz$ & $oe$ \\
    \hline\hline
    \end{tabular}}
\end{table}

Putting them together, for $z=a$ and $C_{2v}$ symmetries, the total wavefunction transforms as $B_1$ or $B_2$ under spin exchange, and identical bosons transform as $A_1$ or $A_2$, giving us the state chart that constrains the allowed coupled ($\Gamma_{\text{nuc}} \otimes \Gamma_{\text{rot}}$) system angular momentum in Table~\ref{tab:bosferm}.

\begin{table}[h!]
\centering
\caption{Allowed combinations of rotational and nuclear-spin irreps for bosons and fermions under $C_{2v}$.}
\label{tab:bosferm}
\begin{tabular}{ccc|cc|cc}
\hline\hline
\multicolumn{3}{c|}{Rotational} & \multicolumn{2}{c|}{Bosons} & \multicolumn{2}{c}{Fermions} \\ \hline
$(-1)^{K_a}$ & $(-1)^{K_c}$ & $\Gamma_{\text{rot}}$ & $\Gamma_{\text{nuc}}$ & $\Gamma_{\text{tot}}$ & $\Gamma_{\text{nuc}}$ & $\Gamma_{\text{tot}}$ \\ \hline
$+$ & $+$ & $A_1$ & $A_1$ & $A_1$ & $B_2$ & $B_2$ \\ 
$+$ & $-$ & $A_2$ & $A_1$ & $A_2$ & $B_2$ & $B_1$ \\ 
$-$ & $+$ & $B_2$ & $B_2$ & $A_1$ & $A_1$ & $B_2$ \\ 
$-$ & $-$ & $B_1$ & $B_2$ & $A_2$ & $A_1$ & $B_1$ \\ \hline\hline
\end{tabular}
\end{table}

\section{Effective Hamiltonian Matrix Elements}
\label{ap:hamiltonian}

The following are algebraic expressions for the matrix elements employed in the simulation. General angular momentum decoupling can be found in Ref.~\cite{zareAngularMomentumUnderstanding1988}. Throughout, we write the molecular Hamiltonian in the spherical-tensor form and adopt the conventions of Brown and Carrington~\cite{brownRotationalSpectroscopyDiatomic2003}. Each interaction is expressed as a scalar contraction $\sum_k T^k(\mathbf{A})\cdot T^k(\mathbf{B}) = \sum_{k,p}(-1)^p\, T^k_p(\mathbf{A})\,T^k_{-p}(\mathbf{B})$ of rank-$k$ irreducible tensor operators. The rotational kinetic energy is described in the previous section; the spin--rotation interaction is $H_{\rm sr} = \sum_{k} T^k(\bm{\epsilon})\cdot T^k(\mathbf{N},\mathbf{S})$, with $\bm{\epsilon}$ the spin--rotation tensor and $\mathbf{S}$ the electron spin; and the magnetic hyperfine interaction is $H_{\rm hf} = \sum_k a_FT^1(\mathbf{S})\cdot T^1(\mathbf{I})+ T^k(\bm{T})\cdot T^k(\mathbf{S},\mathbf{I})$, with $a_F$ the Fermi contact term, $\bm{T}$ the electron-spin--nuclear-spin dipolar coupling tensor and $\mathbf{I}$ the nuclear spin. Tensor components are referred to the molecule-fixed axis $\hat{a}$ (taken as the $a$ principal axis, so that $K \equiv K_a$) and the laboratory axis $\hat{Z}$, related by the Wigner rotation matrices $\mathcal{D}^{(k)}_{pq}(\omega)$; $K$ and $M$ denote the projections of $\mathbf{N}$ on $\hat{a}$ and of $\mathbf{F}$ on $\hat{Z}$, respectively.

The $P,T$-odd energy shift of a polarized open-shell molecule receives contributions from both the \mbox{(free-)electron} EDM $d_e$ and the $P,T$-odd scalar--pseudoscalar nucleon--electron coupling constant $k_s$; the two are indistinguishable within any single system~\cite{Pospelov2005,ChuppElectric2019}. We write the interaction as
\begin{equation}
    H_{\text{EDM}} = \left(W_d\, d_e + W_s\, k_s\right)\,\mathbf{S}\cdot\hat{a},
\end{equation}
where the eEDM enhancement factor $W_d$ and the coupling constant $W_s$ encode the relativistic electronic structure near the heavy nucleus. For $^2\Sigma$ electronic states, the effective electric field is conventionally defined as $\mathcal{E}_{\text{eff}} \equiv -W_d/2$, where $\mathcal{E}_{\text{eff}}$ is a signed quantity, with positive values corresponding to the effective electric field pointing in the same direction as the molecular-frame EDM, $\hat{a}$, and negative values otherwise. It reaches tens of GV/cm in heavy-metal species---four to five orders of magnitude larger than attainable laboratory fields. The eEDM interaction is then written as $-d_e \, \mathcal{E}_{\text{eff}} \, 2 \, \langle\mathbf{S}\cdot\hat{a}\rangle$, so that the polarization factor is carried by the operator; in the fully polarized limit, $\langle\mathbf{S}\cdot\hat{a}\rangle=1/2$ . Since a single measurement bounds only the combination of the two couplings, results are conventionally quoted on the equivalent electron EDM,
\begin{equation}
    d_e^{\rm equiv} \equiv d_e + \frac{W_s}{W_d}\,k_s.
\end{equation}
In Sec.~\ref{sec:eff_ham} we write the interaction in terms of $d_e^{\rm equiv}$, denoted simply $d_e$ for brevity.

The matrix elements take the explicit form:
\begin{widetext}

Rotational term:
\begin{align}
    H_\text{R} 
    &= \bra{N,K,M} |\sum_{k=0,2} T^{k}(B_{\text{rot}}) \cdot T^{k}(\mathbf{N},\mathbf{N})| N',K',M'\ket \nonumber\\
    &= \sum_{k=0,2} \sum_{q}
       \delta_{N,N'} \delta_{M,M'}
       (-1)^{N-M}
       \threeJ{N}{-M}{0}{0}{N}{M}
       \sixJ{k}{k}{0}{N}{N}{N}
       (-1)^{N-K} \nonumber\\
    &\quad \times
       (2N+1)\,
       \threeJ{N}{-K}{k}{q}{N'}{K'}
       \sqrt{2k+1}\,
       \sixJ{1}{1}{k}{N}{N}{N}\,
       N(N+1)(2N+1)\, T^{k}_{q}(B)\,,
\end{align}
with
\begin{align}
    T^0_0(B) &= -\frac{1}{\sqrt{3}}(A+B+C),\\
    T^2_0(B) &= \frac{1}{\sqrt{6}}(2A-B-C),\\
    T^2_2(B) &= T^2_{-2}(B) = \frac{1}{2}(B-C).
\end{align}

Spin-rotation term:
\begin{align}
    H_\text{SR}
    &= \bra{N,K;S,J,I_N,I_H,F,M}
       |\frac{1}{2} \sum_{k=0,2}
       \big[ T^{k}(\bm{\epsilon})\cdot T^{k}(\mathbf{N},\mathbf{S}) 
           + T^{k}(\mathbf{N},\mathbf{S})\cdot T^{k}(\bm{\epsilon}) \big]|
       N',K';S,J',I_N,I_H,F,M\ket \nonumber\\
    &= \sum_{k=0,2} \sum_{q}
       \sqrt{2k+1}\,
       (-1)^{N+S+J}
       \sixJ{N'}{S}{J}{S}{N}{1}
       \sqrt{S(S+1)(2S+1)}
       \sqrt{(2N+1)(2N'+1)} \nonumber\\
    &\quad \times
       \frac{1}{2}
       \Big[
         \sixJ{k}{1}{1}{N'}{N}{N'} 
         \sqrt{N'(N'+1)(2N'+1)}
         +
         \sixJ{k}{1}{1}{N}{N'}{N}
         \sqrt{N(N+1)(2N+1)}
       \Big] \nonumber\\
    &\quad \times
       (-1)^{N-K}
       \threeJ{N}{-K}{k}{q}{N'}{K'}
       T^{k}_{q}(\epsilon)\,
       \delta_{F,F'}\,\delta_{M,M'}\,\delta_{J,J'}\,,
\end{align}
with
\begin{align}
    T^0_0(\epsilon) &= -\frac{1}{\sqrt{3}}(\epsilon_{aa}+\epsilon_{bb}+\epsilon_{cc}),\\
    T^2_0(\epsilon) &= \frac{1}{\sqrt{6}}(2\epsilon_{aa}-\epsilon_{cc}-\epsilon_{bb}),\\
    T^2_2(\epsilon) &= T^2_{-2}(\epsilon) = \frac{1}{2}(\epsilon_{bb}-\epsilon_{cc}).
\end{align}

Hydrogen hyperfine:
\begin{align}
H_{\text{HFS}}^{(H)} 
& =\bra N,K;S,J,I_N,F_N,I_H,F,M_F|
  \sum_{i} a_{F}^{(H)} T^1(\mathbf{I}_H) \cdot T^1(\mathbf{S}) + T^2(\mathbf{T_{I_\textit{H}}})\cdot  T^2(\mathbf{I}_H,\mathbf{S})|
 N',K';S,J',I_N,F_N',I_H,F',M_F'\ket
\nonumber\\[4pt]
&=
\delta_{F F'}\,\delta_{M_F M_F'}\,
(-1)^{F_N' + I_H + F}
\sqrt{S(S+1)(2S+1)\;
      I_H(I_H+1)(2I_H+1)\;
      (2J+1)(2J'+1)}
\nonumber\\
&\quad\times
\sixJ{I_H}{F_N'}{F}{F_N}{I_H}{1}\,
\sqrt{(2F_N+1)(2F_N'+1)}\,
(-1)^{F_N' + J + I_N + 1}\,
\sixJ{J'}{F_N'}{I_N}{F_N}{J}{1}
\nonumber\\[4pt]
&\quad\times
\Bigg[
\delta_{N N'}\,\delta_{K K'}\,
(-1)^{J+S+N+1}\,
\sixJ{S}{J'}{N}{J}{S}{1}\,a_H
\nonumber\\
&\qquad
-\sqrt{30}\,\sqrt{(2N+1)(2N'+1)}\,
\nineJ{N}{N'}{2}{S}{S}{1}{J}{J'}{1}\,
(-1)^{N-K}
\nonumber\\
&\qquad\quad\times
\Bigg\{
\threeJ{N}{-K}{2}{0}{N'}{K'}\,
\frac{T^{(H)}_{aa}}{2}
+
\left[
\threeJ{N}{-K}{2}{2}{N'}{K'}
+
\threeJ{N}{-K}{2}{-2}{N'}{K'}
\right]
\sqrt{\frac{1}{24}}\,
\bigl(T^{(H)}_{aa}+2T^{(H)}_{bb}\bigr)
\Bigg\}
\Bigg].
\end{align}

Nitrogen hyperfine:
\begin{align}
H_{\text{HFS}}^{(N)} 
& = \bra N,K;S,J,I_N,F_N,I_H,F,M_F|
 \sum_{i} a_{F}^{(N)} T^1(\mathbf{I}_N) \cdot T^1(\mathbf{S}) + T^2(\mathbf{T_{I_\textit{N}}})\cdot  T^2(\mathbf{I}_N,\mathbf{S})|N',K';S,J',I_N,F_N',I_H,F',M_F'\ket
\nonumber\\[4pt]
&=
\delta_{F F'}\,\delta_{M_F M_F'}\,\delta_{F_N F_N'}\,
(-1)^{J' + I_N + F_N}
\sqrt{S(S+1)(2S+1)\;
      I_N(I_N+1)(2I_N+1)\;
      (2J+1)(2J'+1)}
\nonumber\\
&\quad\times
\sixJ{I_N}{J'}{F_N}{J}{I_N}{1}
\nonumber\\[4pt]
&\quad\times
\Bigg[
\delta_{N N'}\,\delta_{K K'}\,
(-1)^{J+S+N+1}\,
\sixJ{S}{J'}{N}{J}{S}{1}\,a_N
\nonumber\\
&\qquad
-\sqrt{30}\,\sqrt{(2N+1)(2N'+1)}\,
\nineJ{N}{N'}{2}{S}{S}{1}{J}{J'}{1}\,
(-1)^{N-K}
\nonumber\\
&\qquad\quad\times
\Bigg\{
\threeJ{N}{-K}{2}{0}{N'}{K'}\,
\frac{T^{(N)}_{aa}}{2}
+
\left[
\threeJ{N}{-K}{2}{2}{N'}{K'}
+
\threeJ{N}{-K}{2}{-2}{N'}{K'}
\right]
\sqrt{\frac{1}{24}}\,
\bigl(T^{(N)}_{aa}+2T^{(N)}_{bb}\bigr)
\Bigg\}
\Bigg].
\end{align}

Stark term:
\begin{align}
    H_{\mathcal{E}}
    &= \bra{N,K;S,J,I_N,F_N,I_H,F,M}
       |-T^{1}(\mathbf{D_0})\cdot T^{1}(\bm{\mathcal{E}})|
       N',K';S,J',I_N,F_N',I_H,F',M'\ket \nonumber\\
    &= (-1)^{F-M+1}
       \threeJ{F}{-M}{1}{0}{F'}{M'}
       (-1)^{F'+F_N+I_H+1}
       \sqrt{(2F+1)(2F'+1)}
       \sixJ{F_N'}{F'}{I_H}{F}{F_N}{1} \nonumber\\
    &\quad \times
       (-1)^{F_N'+J+I_N+1}
       \sqrt{(2F_N+1)(2F_N'+1)}
       \sixJ{J'}{F_N'}{I_N}{F_N}{J}{1} \nonumber\\
    &\quad \times
       (-1)^{J'+N+S+1}
       \sqrt{(2J+1)(2J'+1)}
       \sixJ{N'}{J'}{S}{J}{N}{1} \nonumber\\
    &\quad \times
       \mathcal{E}_Z D_0\,
       (-1)^{N-K}
       \sqrt{(2N+1)(2N'+1)}
       \threeJ{N}{-K}{1}{0}{N'}{K'}\,.
\end{align}

Isotropic Zeeman term, where $\mu_B$ is the Bohr magneton and $g_S$ is the free-electron $g$-factor:
\begin{align}
    H_{\mathcal{B}}
    &= \bra{N,K;S,J,I_N,F_N,I_H,F,M}
       |T^{1}(\mathbf{S})\, \cdot T^1(\bm{\mathcal{B}}) \mu_B g_S|
       N',K';S,J',I_N,F_N',I_H,F',M'\ket \nonumber\\
    &= \delta_{NN'}\delta_{KK'}\mathcal{B}_Z \mu_B g_S\,
       (-1)^{F-M+F'+F_N+I_H+1+F_N'+J+I_N+1+J+S+N+1}
       \threeJ{F}{-M}{1}{0}{F'}{M'}
       \sqrt{(2F+1)(2F'+1)} \nonumber\\
       &\times \sqrt{(2F_N+1)(2F_N'+1)} 
       \sqrt{(2J+1)(2J'+1)}
       \sqrt{(2S+1)\,S(S+1)}\nonumber\\
       & \times\sixJ{F_N'}{F'}{I_H}{F}{F_N}{1}
       \sixJ{J'}{F_N'}{I_N}{F_N}{J}{1}
       \sixJ{S}{J'}{N}{J}{S}{1}\,.
\end{align}

For anisotropic (Curl) Zeeman shift, we have:
\begin{align}
H_{\text{Curl}}
    &= \bra{N,K;S,J,I_N,F_N,I_H,F,M}
       |\sum_{k=0,2}\sqrt{\frac{1}{3}(2k+1)}T^1(\tilde {\mathbf{g}}^k,\mathbf{S})\cdot T^1(\bm{\mathcal{B}})\mu_B|
    N',K';S,J',I_N,F_N',I_H,F',M'\ket \nonumber\\
    & = \sum_{k=0,2} \sum_q (-1)^{F-M+F'+F_N+I_H+1+F_N'+J+I_N+1}\threeJ{F}{-M}{1}{0}{F'}{M'}
       \sqrt{(2F+1)(2F'+1)} \nonumber\\
       & \times\sqrt{(2F_N+1)(2F_N'+1)} \sixJ{F_N'}{F'}{I_H}{F}{F_N}{1}\;\sixJ{J'}{F_N'}{I_N}{F_N}{J}{1} (-1)^k\sqrt{(2k+1)}\sqrt{(2J+1)(2J'+1)}\nineJ{N}{N'}{k}{S}{S}{1}{J}{J'}{1}\nonumber \\
       & \times \sqrt{S(S+1)(2S+1)}(-1)^{N-K}\sqrt{(2N+1)(2N'+1)}\threeJ{N}{-K}{k}{q}{N'}{K'}T^k_q(\tilde{\mathbf{g}})\mathcal{B}_Z\mu_B,    
\end{align}
where $T^k_q(\tilde{\mathbf{g}})$ is expressed similarly compared to spin-rotation tensor $T^k_q(\epsilon)$.

Nitrogen Zeeman term, where $\mu_N$ is the nuclear magneton and $g_N$ is the nitrogen nuclear $g$-factor:
\begin{align}
H_{\mathcal{B},\text{N}}
    &= \bra{N,K;S,J,I_N,F_N,I_H,F,M}|
       \, T^1(\mathbf{I}_N)\cdot T^1(\bm{\mathcal{B}}) \mu_N g_N| \,
       N',K';S,J',I_N,F_N',I_H,F',M'\ket \nonumber\\
    &= (-1)^{F-M+F'+F_N+I_H+1+F_N'+J+I_N+1}\,
       \threeJ{F}{-M}{1}{0}{F'}{M'}\,
       \sqrt{(2F+1)(2F'+1)} \nonumber\\
    &\quad \times \sqrt{(2F_N+1)(2F_N'+1)}\,
       \sixJ{F_N'}{F'}{I_H}{F}{F_N}{1}\;
       \sixJ{I_N}{F_N'}{J}{F_N}{I_N}{1} \nonumber\\
    &\quad \times \delta_{NN'}\,\delta_{KK'}\,\delta_{JJ'}\,
       \sqrt{I_N(I_N+1)(2I_N+1)}\mathcal{B}_Z\mu_Ng_N\,.
\end{align}

Hydrogen Zeeman term, where $g_H$ is the hydrogen nuclear $g$-factor:
\begin{align}
    H_{\mathcal{B},\text{H}}
    &= \bra{N,K;S,J,I_N,F_N,I_H,F,M}
       |T^{1}(\mathbf{I}_H)\, \cdot T^1(\bm{\mathcal{B}})\, \mu_N g_H|
       N',K';S,J',I_N,F_N',I_H,F',M'\ket \nonumber\\
    &= \delta_{NN'}\delta_{KK'}\delta_{JJ'}\delta_{F_N F_N'}\,
       \mathcal{B}_Z\, \mu_N g_H\,
       (-1)^{F-M+F'+F_N+I_H+1}
       \threeJ{F}{-M}{1}{0}{F'}{M'}
       \sqrt{(2F+1)(2F'+1)} \nonumber\\
       &\times \sqrt{I_H(I_H+1)(2I_H+1)}\,
       \sixJ{I_H}{F'}{F_N}{F}{I_H}{1}\,.
\end{align}

For Nitrogen spin $\geq 1$, such as $^{14}\text{N}$, the electric quadrupole moment (EQM) term manifests as
\begin{align}
H_Q
&=
\bra{N,K;S,J,I_N,F_N,I_H,F,M}
  |eT^2(\mathbf{Q}) \cdot T^2(\nabla \bm{\mathcal{E}}_{\text{internal}})|
N',K';S,J',I_N,F_N',I_H,F',M'\ket
\nonumber\\[4pt]
&=
\delta_{F F'}\,\delta_{M M'}\,
\left[\threeJ{I_N}{-I_N}{2}{0}{I_N}{I_N}\right]^{-1}
(-1)^{J' + I_N + F_N}\,
(-1)^{N + S + J'}
\nonumber\\[4pt]
&\quad\times
\sixJ{I_N}{J}{F_N}{J'}{I_N}{2}\,
\sqrt{(2J'+1)(2J+1)}\,
\sqrt{(2N'+1)(2N+1)}\,
\sixJ{N'}{J'}{S}{J}{N}{2}
\nonumber\\[4pt]
&\quad\times
\Bigg\{
(-1)^{N-K}\,
\threeJ{N}{-K}{2}{0}{N'}{K'}\,
\frac{\chi_{aa}}{4}
\nonumber\\[2pt]
&\qquad\qquad\qquad
+
\left[
(-1)^{N-K}\threeJ{N}{-K}{2}{2}{N'}{K'}
+
(-1)^{N-K}\threeJ{N}{-K}{2}{-2}{N'}{K'}
\right]
\frac{2\chi_{bb}+\chi_{aa}}{4\sqrt{6}}
\Bigg\}.
\end{align}

EDM term:
\begin{align}
    H_{\text{EDM}}
    &= \bra{N,K;S,J,I_N,F_N,I_H,F,M}|
       T^{1}_{q=0}(\mathbf{S})|
       {N',K';S,J',I_N,F_N',I_H,F',M'}\ket \nonumber\\
    &= \delta_{F,F'}\,\delta_{M,M'}\,\delta_{F_N,F_N'}\,\delta_{J,J'}\
       (-1)^{N'+S+J}
       \sixJ{S}{N'}{J}{N}{S}{1} \nonumber\\
    &\quad \times
       (-1)^{N-K}
       \sqrt{(2N+1)(2N'+1)}
       \threeJ{N}{-K}{1}{0}{N'}{K'}
       \sqrt{S(S+1)(2S+1)}\,.
\end{align}

\end{widetext}

The Stark and Zeeman matrix elements above assume the field lies along the laboratory quantization axis $\hat{Z}$. A field of arbitrary orientation is incorporated by decomposing it into lab-frame rank-1 spherical components and summing over them. Taking the Stark interaction as the example (the Zeeman case follows identically), the general Hamiltonian is
\begin{equation}
    H'_{\mathcal{E}} = -\sum_{p = 0,\pm 1}(-1)^{p}\,T^1_{p}(\mathbf{D})\,
    T^1_{-p}(\bm{\mathcal{E}}),
\end{equation}
\begin{equation}
    \mathcal{E}_{0} = \mathcal{E}_{Z},\quad
    \mathcal{E}_{\pm 1} = \mp\tfrac{1}{\sqrt{2}}\bigl(\mathcal{E}_{X} \pm i\mathcal{E}_{Y}\bigr).
\end{equation}
Relative to the longitudinal case, only the lab-frame $3j$ symbol from the
Wigner--Eckart reduction of $T^1_p(\mathbf{D})$ is modified,
\begin{equation}
    \begin{pmatrix} F & 1 & F' \\ -M & 0 & M' \end{pmatrix}
    \;\longrightarrow\;
    \begin{pmatrix} F & 1 & F' \\ -M & p & M' \end{pmatrix},
\end{equation}
with the selection rule $M' = M - p$ enforced by the symbol: the transverse
components $p=\pm1$ couple states with $\Delta M = \mp 1$, while all
molecule-frame factors are unchanged. The transverse field therefore breaks the
block-diagonality in $M$, and the Hamiltonian must be diagonalized over the
full $M$ manifold rather than within fixed-$M$ blocks.

\begin{table*}[t]
    \centering
    \caption{Comparison of the reference and refit $r_0$ geometries together with the corresponding rotational constants for $\mathcal{M}$--NH$_2$ ($\mathcal{M}$~=~Ca, Sr). Reference geometries and experimentally determined rotational constants are taken from Refs.~\cite{Brewsterpure2000,SheridanRotational2005a}. Primed rotational constants corresponding to CaND$_2$ were also used in the structural fitting.}
    \label{tab:refit}    \begin{ruledtabular}
    \begin{tabular}{llllllllllll}
         &  Geometry &  $r_\text{$\mathcal{M}$--N}$& $r_\text{H--N}$& $\alpha_\text{$\mathcal{M}$--N--H}$ &Rot.~const.& $A$& $B$& $C$& $A$'& $B$'&$C$'\\
         \hline
         CaNH$_2$&  reference&  2.126& 1.018& 127.1 &$\rightarrow$ calc. & 380324& 8987.8& 8780.3& 190308& 7788.8&7482.5\\
         &  refit&  2.126& 1.010& 127.6 &$\rightarrow$ calc. & 391564& 8997.0& 8794.9& 195933& 7795.3&7497.0\\
 & & & &  &experiment& 392127& 9009.1& 8782.8& 195669& 7807.6&7484.7\\
 SrNH$_2$& reference& 2.256& 1.021& 127.25 &$\rightarrow$ calc. & 379597& 6788.4& 6669.1& & &\\
 & refit& 2.257& 1.007& 127.6 &$\rightarrow$ calc. & 394345& 6782.3& 6667.6& & &\\
 & & & &  &experiment& 394353& 6790.3& 6659.5& & &\\
    \end{tabular}
    \end{ruledtabular}
\end{table*}

\begin{table*}[t]
    \centering
    \caption{Comparison of calculated and reference geometries for $\mathcal{M}$--NH$_2$ species.}
    \label{tab:geom}    \begin{ruledtabular}
    \begin{tabular}{lllllll}
         & Type & $r_\text{$\mathcal{M}$--N}$& $r_\text{H--N}$& $\alpha_\text{$\mathcal{M}$--N--H}$ &$\alpha_\text{H--N--H}$ &Reference\\
         \hline
         CaNH$_2$&  $R_e$ calc.& 2.126& 1.015& 127.4 & 105.3 &\\
                 & $R_0$ ref. &  2.126& 1.010& 127.6 & 104.8 & \cite{Brewsterpure2000} (refit)\\
         SrNH$_2$& $R_e$ calc.&  2.263& 1.016& 127.6 & 104.8 &\\
                 & $R_0$ ref. &  2.257& 1.007& 127.6 & 104.8 & \cite{SheridanRotational2005a} (refit)\\
         BaNH$_2$& $R_e$ calc.&  2.380& 1.017& 127.5 & 105.0 &\\
         RaNH$_2$& $R_e$ calc.&  2.476& 1.017& 127.6 & 104.7 &\\
                 & $R_e$ ref. &  2.431& 1.016& 127.8 & 104.4 & \cite{zhangAnalyticGradientsRelativistic2023}\\
    \end{tabular}
    \end{ruledtabular}
\end{table*}

\section{Molecular Geometries}\label{ap:refit}

During this study, it was realized that the structural fits of the CaNH$_2$ and SrNH$_2$ geometries given in the original works~\cite{Brewsterpure2000,SheridanRotational2005a} were showing large deviations from the measured rotational constants. This is due to the numerical fitting procedure based on fitting the geometry parameters directly to the total moments of inertia. 
Due to the near-linear nature of the $\mathcal{M}$--NH$_2$ molecules with only the light hydrogen atoms off the central axis, there is a factor of about 50 difference between the individual rotational constants. This effectively gives the $I_b$ and $I_c$ components $\sim50\times$ larger weight in the fitting procedure compared to $I_a$, leading to a rather poor fit for the rotational constant $A\sim1/I_a$. 

We have refit the geometry using relative residuals instead of fitting to the absolute moments of inertia, leading to a more balanced fit and a much closer agreement with the measured rotational constants. The refit geometries are compared to the previously derived structures in Table~\ref{tab:refit} together with the corresponding rotational constants calculated using the rigid-rotor approximation and compared to the experimentally determined rotational constants. 
Most notably, the resulting N--H bond length is 0.008~\AA~and 0.014~\AA~shorter for CaNH$_2$ and SrNH$_2$, respectively, compared to the previously derived geometries. 
Note that using relative residuals for rotational constants instead of moments of inertia (or a combination of both) in the fit produces near-identical results with agreement in 5--6 significant figures.

In the case of CaNH$_2$, simultaneous fitting with the rotational constants of the heavier isotopologue CaND$_2$ allows for extra degrees of freedom. Although the structure was refit with the dihedral angle as a free parameter, the best fit is fully planar. In the case of SrNH$_2$, only 3 free parameters are available for the fit. Nevertheless, when either of the parameters shown in Table~\ref{tab:refit} was kept constant while the dihedral angle was relaxed, the fitting consistently results in a planar structure.

Finally, in Table~\ref{tab:geom}, we provide a comparison of the geometries calculated using quasi-relativistic DFT using the B3LYP functional~\cite{stephensInitioCalculationVibrational1994} and the Def2-TZVPP basis set~\cite{kauppPseudopotentialApproachesCa1991,weigendBalancedBasisSets2005} with the available reference geometries for CaNH$_2$, SrNH$_2$ and RaNH$_2$ serving as validation for the use of the calculated geometry for BaNH$_2$, for which no reference was found.
Reference geometries for CaNH$_2$ and SrNH$_2$ are based on the above refit of the available experimental data~\cite{Brewsterpure2000,SheridanRotational2005a,ThompsenRotational2000}. 
For RaNH$_2$, the available high-level theoretical geometry was used~\cite{zhangAnalyticGradientsRelativistic2023} (note the misprint in Ref.~\cite{zhangAnalyticGradientsRelativistic2023}; the H--N--H angle is actually the Ra--N--H angle).

\bibliography{references}

\end{document}